\documentclass[journal ]{new-aiaa}
\usepackage[utf8]{inputenc}
\usepackage{textcomp}

\usepackage{subcaption}
\usepackage{hyperref}
\usepackage{float}
\usepackage{empheq}
\usepackage{color, colortbl}
\usepackage{caption}
\usepackage{wrapfig}
\usepackage{footmisc}
\usepackage{graphicx}
\usepackage{amsmath}
\usepackage[version=4]{mhchem}
\usepackage{siunitx}
\usepackage{longtable,tabularx}

\usepackage{multicol}
\usepackage{nomencl}
\makenomenclature
\usepackage{etoolbox}
\renewcommand\nomgroup[1]{%
  \item[\bfseries
  \ifstrequal{#1}{S}{Symbols}{%
  \ifstrequal{#1}{A}{Abbreviations}{%
  }}%
]}

\def\Onera{ONERA}
\newcommand{\Xiv}{{\boldsymbol\Xi}}
\newcommand{\xiv}{{\boldsymbol\xi}}
\newcommand{\Nset}{\mathbb{N}}
\newcommand{\Rset}{\mathbb{R}}
\newcommand{\esp}{\mathbb{E}}
\newcommand{\iexp}{\mathrm{e}}
\newcommand{\itr}{\textsf{T}}
\newcommand{\norm}[1]{|#1|}
\newcommand{\nord}[1]{\|#1\|}
\newcommand{\inner}[1]{\left\langle#1\right\rangle}
\newcommand{\demi}{\frac{1}{2}}
\newcommand{\eref}[1]{Eq.~(\ref{#1})}
\newcommand{\sref}[1]{Sect.~\ref{#1}}
\newcommand{\fref}[1]{Fig.~\ref{#1}}
\newcommand{\Fref}[1]{Fig.~\ref{#1}}
\newcommand{\tref}[1]{Table~\ref{#1}}

\newcommand{\iin}{\text{in}}
\newcommand{\its}{\text{ts}}
\newcommand{\porder}{K}
\newcommand{\torder}{\kappa}
\newcommand{\pdf}{\pi}
\newcommand{\NLaw}{{\mathcal N}}
\newcommand{\npdf}{n}
\newcommand{\CV}{CV}
\newcommand{\RC}{R_C}
\newcommand{\RD}{R_{3D/2D}}
\newcommand{\velocity}{\boldsymbol{u}}
\newcommand{\meant}[1]{\overline{#1}}
\newcommand{\favre}[1]{\widetilde{#1}}
\newcommand{\enthalpy}{h}
\newcommand{\Reynolds}{\text{Re}}
\newcommand{\QoI}{\vartheta}

\newcommand{\Mach}{M}

\newcommand{\Machts}{\Mach_\text{ts}}
\newcommand{\marie}[1]{\textcolor{black}{#1}}
\newcommand{\alert}[1]{\textcolor{black}{#1}}
\newcommand{\nicola}[1]{\textcolor{black}{#1}}

\newcommand{\bibrev}[1]{\textcolor{black}{#1}}

\title{Uncertainty Quantification Applied to the Propagation of a Transonic Wind Tunnel Inflow Inhomogeneities}

\newcommand*\samethanks[1][\value{footnote}]{\footnotemark[#1]}
\author{Nicola Detomaso \footnote{Graduate student, Department of Mechanics, Mathematics and Management.}}
\affil{Politecnico di Bari, 70126 Bari, Italy}
\author{Vincent Brion\footnote{Research Engineer, Department of Aerodynamics, Aeroelasticity and Acoustics (DAAA).}, Julien Dandois\samethanks, Marie Couliou\samethanks}
\affil{\Onera, Paris-Saclay University, 92190 Meudon, France}
\author{\'Eric Savin \footnote{Research Engineer, Department of Information Processing and Systems (DTIS).}}
\affil{\Onera, Paris-Saclay University, 91120 Palaiseau, France}

\begin{document}

\maketitle

\begin{abstract}

The uncertainty associated with the experimental inflow in a wind tunnel affects the prediction of the flow of interest by numerical simulations. We evaluate this impact using uncertainty quantification. A method is developed and applied to the simulation of the drag generated by the flow past a cylinder installed in the transonic S3Ch ONERA mid-scale facility. The inflow uncertainty results from the imperfect knowledge and variability of the flow in the settling chamber. It is taken into account via the inlet boundary condition in the numerical companion setup and evaluated experimentally by measuring the inflow using a hot-wire rake. The propagation of the input uncertainties is carried \alert{out} through a two-dimensional RANS model of the experiment. A polynomial surrogate model is developed to infer the uncertainty associated with the drag of the cylinder. Following observations of Gaussian inputs, the parameters of the stochastic model are constructed in two ways, first through a projection approach, based on the Gauss-Hermite quadrature rule, and then using a sparsity based regression approach, based on compressed sensing. The latter drastically reduces the number of deterministic numerical simulations. The drag is most influenced by the central part of the inflow but the overall uncertainty remains low.
\end{abstract}

\newpage

\nomenclature[S]{$\varnothing$}{diameter}
\nomenclature[S]{$c_j$}{polynomial chaos expansion coefficient}
\nomenclature[S]{$C_p$}{pressure coefficient}
\nomenclature[S]{$C_D$}{drag coefficient}
\nomenclature[S]{$C_{D,f}$}{\marie{friction drag coefficient}}
\nomenclature[S]{$C_{D,p}$}{\marie{pressure drag coefficient}}
\nomenclature[S]{${\boldsymbol q}$}{\marie{ heat flux}} 
\nomenclature[S]{${\boldsymbol\tau}$}{\marie{viscous stress tensor}}
\nomenclature[S]{$E$}{\marie{total energy}}
\nomenclature[S]{$\rho$}{\nicola{flow density}}
\nomenclature[S]{$\CV$}{\nicola{coefficient of variation}}
\nomenclature[S]{$D$}{number of random variables}
\nomenclature[S]{$\RC$}{Contraction ratio of the nozzle of the wind tunnel}
\nomenclature[S]{$\RD$}{Width ratio between the 3D and 2D nozzles}
\nomenclature[S]{$R$}{cylinder radius}
\nomenclature[S]{$W$, $H$}{settling chamber width, height}
\nomenclature[S]{$w$, $h$, $l$}{test section width, height, length}
\nomenclature[S]{$\Mach^\circ$}{inflow Mach number in the settling chamber reconstructed from experiment}
\nomenclature[S]{$\Mach_\its$}{Mach number in the test section}
\nomenclature[S]{$u_\its$}{freestream velocity in the test section}
\nomenclature[S]{$\Reynolds$}{ Reynolds number}
\nomenclature[S]{$N$}{number of random samples}
\nomenclature[S]{$p$}{\marie{ pressure}}
\nomenclature[S]{$\porder$}{number of expansion coefficients}
\nomenclature[S]{$\torder$}{polynomials total order}
\nomenclature[S]{$Q$}{number of quadrature nodes}
\nomenclature[S]{$\npdf$}{\nicola{Gaussian distributions}}
\nomenclature[S]{$w_i$}{quadrature weights}
\nomenclature[S]{$\mu$}{average}
\nomenclature[S]{$\psi_j$}{one-dimensional orthonormal polynomial of degree $j$}
\nomenclature[S]{$\Psi_{\bf j}$}{multi-dimensional orthonormal polynomial of degrees ${\bf j}$}
\nomenclature[S]{$\pdf$}{\nicola{probability density function of the random input variables}}
\nomenclature[S]{$\sigma$}{standard deviation}
\nomenclature[S]{$\QoI$}{quantity of interest}
\nomenclature[S]{$\Xiv$}{random input variables}
\nomenclature[S]{$g_\porder$}{\nicola{polynomial surrogate model}}
\nomenclature[S]{$S_d$}{\marie{Solob's indice}}
\nomenclature[S]{$\xiv_i$}{quadrature node/sample of the \nicola{numerical} random input variables}
\nomenclature[S]{$\velocity$}{\nicola{velocity vector}}
\nomenclature[S]{$P_\iin$}{\nicola{inflow stagnation pressure}}
\nomenclature[S]{$T_\iin$}{\nicola{inflow stagnation temperature}}
\nomenclature[S]{$\enthalpy$}{\marie{specific enthalpy}}
\nomenclature[S]{$H_\iin$}{\nicola{inflow stagnation enthalpy}}
\nomenclature[S]{$c_v$, $c_p$}{\nicola{isochoric and isobaric air specific heat capacity}}

\nomenclature[A]{CFD}{Computational Fluid Dynamics}
\nomenclature[A]{CS}{Compressed Sensing}
\nomenclature[A]{GQ}{Gauss quadrature}
\nomenclature[A]{KL}{Kullback-Leibler}
\nomenclature[A]{PCE}{Polynomial Chaos Expansion}
\nomenclature[A]{PDF}{Probability Density Function}
\nomenclature[A]{QoI}{Quantity of Interest}
\nomenclature[A]{RANS}{\alert{Reynolds-Averaged Navier-Stokes}}
\nomenclature[A]{UQ}{Uncertainty Quantification}

\begin{multicols}{2}
\printnomenclature
\end{multicols}

\section{Introduction}
\label{Introduction}

The validation of \bibrev{computational fluid dynamics (CFD)} tools using wind tunnel experiments poses many questions on the way to setup the test case and on the necessary test data to use in the numerical configuration to make the numerical simulation most compatible to the real flow~\cite{barlow1999low,goethert1961transonic,marvin1987wind,bradley1988cfd}.
In this process the account of real information on the flow generated by the wind tunnel appears as an important matter often neglected in the simulation work. In fact a wind tunnel represents a complex system \cite{cattafesta2010fundamentals} that generates testing conditions not necessarily well-known. This leads to an intrinsic uncertainty in the proper setting and initialisation of the numerical model to target experimental reproduction that, on general grounds, includes geometrical and flow parameters alike. As already noticed~\cite{owen2008measurement,steinle1982wind} the flow environment in wind tunnel testing is rarely properly characterized or documented while the quality of the flow (turbulence level, uniformity, steadiness) can have a profound impact on the aerodynamics of the experiments and the output results~\cite{manshadi2011importance}. The qualification of the flow quality is done occasionally~\cite{allen2014qualification} or when required by an upgrade of the facility~\cite{krynytzky2002boeing,szoke2020developing,vishwanathan2020aerodynamic} to check the improvements. Yet regular qualifications of the flow are difficult in practice as it requires time and much dedicated work~\cite{ljungskog2019uncertainty}, and therefore unreported variations from nominal properties are to be expected. Furthermore the geometry of a model or of the test circuit can also suffer from departure from initial design for various reasons (quality of manufacturing, manual adjustment of mechanical parts, presence of clearance between parts, thermal deformations, etc.). Following these comments, the reproduction of experimental results by numerical simulations naturally faces the question of whether these discrepancies generally discarded by ignorance or feasibility of regular assessments have an effect on the final numerical \alert{results}.

This topical question of the effect of the experimental environment on the result of the numerical calculations or the sensitivity of the latter can be tackled by considering a stochastic framework and looking at the testing environment, to be reproduced by the simulation (at least partially), as uncertain. In this work, input uncertainty is used to refer for the lack of knowledge of the inflow of the wind tunnel and its variability~\cite{hoffman1994propagation}. A similar approach is taken by Boon \emph{et al}.~\cite{boon2012reducing}. In their work the effect of the uncertainty in the airfoil model geometry and angle of attack on the aerodynamic forces is explored using a simple panel method of the airfoil aerodynamics and experimental measurements of the uncertain variables. Here the effect of the uncertainty of the incoming flow generated by the wind tunnel is analyzed by assessing the resulting uncertainty associated with the drag of a cylinder placed in the test section, and subjected to a transonic stream. The study is limited to the uncertainty that results from this inhomogeneity and unsteadiness of the inflow into the convergent of the wind tunnel and discards that due to the error in the measurements or that due to the imperfections of the geometry of the test circuit or model. A \alert{Reynolds-Averaged Navier-Stokes (RANS)} model of the flow is used to propagate the uncertainty of the input variables from the inflow frontier to the cylinder aerodynamics.

The presence of uncertainties can significantly restrict the reliability of deterministic computation, shifting the focus to the quantification of the influence of uncertain parameters onto physical systems in order to properly predict the system response to random inputs. One of the most commonly used method for uncertainty quantification (UQ) is the Monte Carlo approach \bibrev{\cite{ROB04}}, but its low convergence rate (\alert{$1/\sqrt{N}$} with $N$ the number of samples) and the consequently large number of samples needed to yield meaningful results make it far too expensive, from a computational point of view, for realistic configurations. Much more efficient UQ methods exist, for instance the \alert{polynomial chaos} (PC) approach \bibrev{\cite{ghanem2003stochastic,soize2004physical,wiener1938homogeneous,xiu2002wiener}}, in its non-intrusive form (NIPC) \bibrev{\cite{le2010spectral}}, which is taken into consideration here. 
The NIPC approach, which does not need code adaptation, is based on performing the deterministic code several times, and selecting samples adequately. \bibrev{Applications in CFD have been considered in \emph{e.g.} \cite{CHA10,DIN10,DOD09,HOS10,KNI06,MAT05,NAJ09,SCH17,SIM10,WEI19,WES17,XIU03}}. If, on one side, the method is suitable for all kind of computational codes, on the other one, it suffers from the so-called "curse of dimensionality" \bibrev{\cite{GIR14}} and, thus, its computational cost increases with the number of random dimensions. For this reason, the \alert{PC} approach is commonly associated to efficient methodologies that investigate the stochastic space of random variables. \alert{In the present work} the generalized \bibrev{PC approach (gPC) \cite{soize2004physical,xiu2002wiener} is complemented with the compressed sensing theory \cite{CAN06,DON06}}, this being compared with the results obtained by \alert{a} Gauss quadrature rule \cite{savin2016sparse}, considered as the reference solution. 
These methods are applied to evaluate the uncertainty associated with the drag of the cylinder due to the uncertainty associated with the inflow conditions. \bibrev{The proposed approach to compute the gPC expansion coefficients by compressed sensing has been implemented in various situations \emph{e.g.} \cite{DOO11,JAK15,MAT12,RUM20,SAL17,tsilifis2019compressive,WES16}; see also \cite{HAM17} for a recent overview and additional references.} The effect of inflow on aerodynamic forces \alert{was recently considered in} \cite{zhang2020effective}, yet generally also \alert{accounting for} uncertainties associated with other parameters (air density and viscosity for instance in \cite{zhang2020effective}). 

A \alert{byproduct} of uncertainty analysis is confidence intervals on data obtained from a combination of experiments and numerical simulations. This represents a valuable information in the objective of performance and robustness. Indeed the level of confidence in a data has consequence on design. Lower confidence leads to larger margins and lower performance while reduced uncertainty allows for bolder optimizations. Typically UQ can be used to optimize a design to lower both some target quantity (for instance the drag of an airfoil) and its standard deviation~\cite{du2019optimum}, thereby targeting efficiency and robustness~\cite{WEI19}. Uncertainty bars on simulation data are important to reinforce the domain of use of aerodynamic design tools~\cite{coleman1997uncertainties,stern2005statistical}. Concepts such as multi-fidelity methods allow to merge information from different approaches including possibly CFD and experiments to deliver the most certain results at a reduced computational cost~\cite{nigam2021toolset,west2017multifidelity,RUM20}. The question of cost and accuracy is central and much research is devoted to improving UQ approach by, for instance, adjusting best polynomial order and the number of sample points~\cite{shimoyama2016uncertainty}, or by identifying, in RANS models, the closure coefficients the most sensitive to the output uncertainty \cite{SCH17}. Mastering uncertainties is also on the path for developing new certification processes inclusive of more simulated data, hence less costly and lengthy. UQ shows some promises in this domain, as shows the works of \cite{june2020system} on the usage of UQ to evaluate uncertainty on noise levels of a flying wing, \alert{or} for certification prediction of the sonic boom based on a reduced set of uncertain parameters \cite{WES17}.
In any way modern CFD, along with wind tunnel methods, faces the challenge of further integration in the certification process~\cite{spalart2016role}. 

The present work, that targets the influence of wind tunnel accuracy is much motivated by this perspective. Specifically the \alert{objective} is to better understand how errors in the values of the freestream produced by the wind tunnel can translate into errors in some target quantity at the model or the flow around it and whether the flow quality related to the inhomogeneity of the incoming flow has a decisive effect on the numerical output. The main interest is on the methodology that we present in details, and the case of the cylinder is taken as an adequate configuration thanks to the availability of data. From an application point of view, there are practical results that are expected from developing such combined numerical and experimental researches that could motivate modifications to the wind tunnel in the future in order, for instance, to improve its design (reducing flow separation by designing adequate pressure loss devices for instance). 

In \alert{\sref{sec:experiments}} we present the configuration of interest upon which the uncertainty quantification is performed. The experimental data \alert{are} described and the numerical model is outlined and validated against the wind tunnel experiment. The methodology for the surrogate model is then introduced \alert{in \sref{sec:UQ}}, along with the compressed sensing analysis. \alert{In \sref{sec:application}} we apply the method to the configuration of interest, propagating the inflow uncertainties into the simulated flow and concluding on the influence on drag prediction. \alert{Finally \sref{sec:CL} offers a summary and conclusions.}

\section{Aerodynamic flow case and stochastic approach}\label{sec:experiments}

\subsection{Experiment}\label{Experiment}

The uncertainty of the inflow in the wind tunnel is associated to the spatial and temporal inhomogeneity of the flow upstream of the test model. These flow defects may result from various sources. The integration of the history of the flow as it passes through the different parts of the wind tunnel circuit (pipes, fan, corners, variations in section size and possibly shape, grids) \alert{causes} secondary flows, multiples wakes, boundary layer and possibly separated flow phenomena that are the primary mechanisms causing turbulence and large scale structures~\cite{moonen2006numerical,idelchik1986handbook}. The resulting turbulence in the settling chamber and downstream test section is seldom changed by modifications of the characteristics of the honeycomb \cite{dryden1931reduction} and the importance of turbulence, both integrated levels and content, on experimental results can be high~\cite{manshadi2011importance}. The overall flow behavior may also depend on the experiment and the particular configuration of the tunnel~\cite{moonen2006numerical,vishwanathan2020aerodynamic}.

\begin{figure}[H]
\centering\includegraphics[width=.6\linewidth]{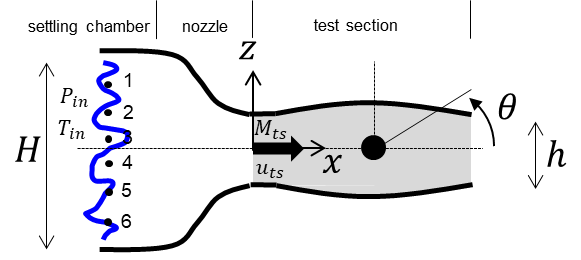}
\caption{Sketch of the configuration of interest illustrating the physical and computational domain for the UQ approach. The test section is highlighted in grey shading. The coordinate system is referenced to the inlet of the test section. The uncertain inflow is characterized by 6 uncertain inputs distributed vertically in the settling chamber and measured by single hot-wire probes installed on a vertical mast. The freestream flow in the test section is qualified by the test section Mach number \textmd{$\Machts$} and \alert{related} velocity \textmd{$u_\its$} obtained from the stagnation conditions \textmd{$P_\iin$}, \textmd{$T_\iin$} and the static pressure in the test section $p_\its$. Note that the exact dimensions are not respected in this plot which serves as an illustration only.  }\label{fig:problem}
\end{figure}

We consider \alert{flows} about a cylinder in the S3Ch transonic wind tunnel of \Onera. The sketch in~\fref{fig:problem} describes the experimental domain made of the settling chamber, the nozzle and the test section of the wind tunnel. The settling chamber has a width $W$ and height $H$ both equal to $4.2m$ and the nozzle is $4m$ long. The test section is $l=2.2m$ long, $w=0.804m$ \alert{wide,} and $h=0.764m$ \alert{high}, yielding a surface ratio between the settling chamber and the test section equal to $\RC=28.7$. The direction of the upstream flow is denoted by $x$, $y$ being the transverse horizontal direction oriented to the right when facing the flow and $z$ the vertical \alert{one}. The reference of axis lies at the entrance of the test section, at mid-height and at the middle of the side walls. The flow velocity is decomposed as $\velocity=\meant{\velocity}+\velocity'$ where $\meant{\cdot}$ denotes time averaging, $\cdot'$ denotes fluctuations of zero mean, and the bold symbol is used for vectors. The velocity vector is $\velocity=\left(u_x,u_y,u_z\right)$. 

The cylinder has a radius $R=20mm$ and lies at $x=1.2m$ from the entrance of the test section. Positions at the cylinder surface are characterized by the angle $\theta \in [-\pi,\pi]$ which is referenced with respect to the positive $x$ direction. The ratio $w/\left(2R \right) \simeq 20 $ ensures a close to two dimensional setup which will be exploited for the numerical simulations to reduce the experimental three-dimensional configuration to a numerical two-dimensional setup. The cylinder is positioned at mid height between the upper and lower wall of the wind tunnel. The upper and lower walls are streamlined in order to reduce wall interference (i.e. perturbation of Mach number and flow incidence from target values are minimized in the region of the model). As sketched in~\fref{fig:problem} wall deformations are symmetric. Note that this wall adaptation is static and accounts solely for the mean flow. The adaptive wall procedure provides a corrected Mach number $\Machts$, equivalent to the Mach number of the same flow in a unconstrained environment, free of walls. Thereafter $\Machts$ is referred to as the test section Mach number. In the present work, $\Machts$ is set to $0.8$. 

The stagnation conditions (pressure $P_\iin$ and temperature $T_\iin$) of the flow are measured in the settling chamber at $x_\iin=-0.328m$ from the entrance of the test section and at a height $z_\iin=0.476H$ and lateral position $y_\iin=-0.469H$. The static pressure $p_\its$ is measured in the test section at $x_\its=0.3m$. It allows to calculate the \alert{test section} Mach number $\Machts$ from \alert{the stagnation pressure} $P_\iin$ using the isentropic flow equations. Together with $\Machts$, using $T_\iin$ we define the \alert{freestream} velocity $u_\its$ to qualify the general velocity of the flow in the test section (note that $u_\its$ is not a directly measured quantity as it is obtained from a computed quantity, the test section Mach number $\Mach_\its$, and stagnation conditions). The typical velocity in the settling chamber is defined as $u_\its/\RC$ and is used to normalize the flow velocity in the settling chamber. The Reynolds number based on the cylinder diameter and freestream velocity $u_\its$ is expressed as $\Reynolds=2Ru_\its/\nu$, and yields a value of $5\times 10^5$ given $\Mach_\its=0.8$ and the stagnation conditions. 

\begin{figure}[H]
\centering
\includegraphics[width=.39\linewidth]{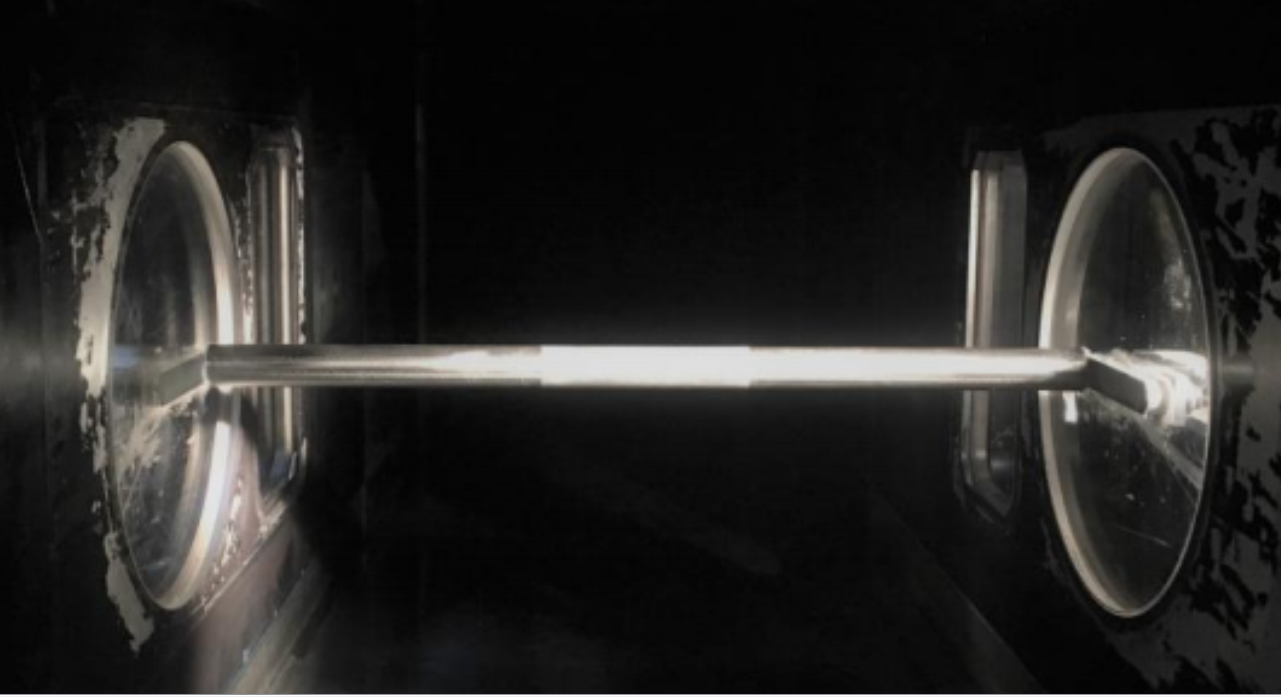}
\caption{Transverse cylinder installed in the test section of the S3Ch wind tunnel. View from upstream. }\label{fig:cylinder}
\end{figure}

The cylinder installed in the test section of the wind tunnel is shown in~\fref{fig:cylinder}, as viewed from upstream. Two large windows at the side enable Schlieren visualizations for flow monitoring, especially the wake pattern. A Phantom 7.3 featuring 800 by 600 pixels and $9.9kHz$ sampling frequency is used to record the Schlieren images. An ensemble of 47 pressure taps, placed at mid-span with an angle \alert{$\theta$} to the $x$ direction, are distributed along the contour of the cylinder to characterize the evolution of the temporal mean of the surface pressure, thereafter expressed in terms of pressure coefficient as \nicola{$C_p=2(p-p_\its)/(\rho_\its u_\its^2)$}, \alert{where $\rho_\its$ is the flow density in the test section (deduced from $\Machts$ and stagnation conditions)}. A PSI\textsuperscript{\textregistered} pressure transducer with 64 ports is used to record the \alert{actual} pressure $p$. 


The integration of this pressure along the surface of the cylinder provides the pressure contribution \marie{$C_{D,p}$} of the total cylinder drag. The cylinder drag coefficient $C_D$ based on the cylinder diameter is obtained from
\begin{equation}
C_D=-\frac{1}{2} \int_{0}^{2\pi} \left( C_p \sin\theta + C_f \cos\theta\right) d\theta= C_{D,p}+C_{D,f}\,.
\end{equation}
The contribution from friction \alert{$C_{D,f}$} can only be obtained from the numerical simulations while the experiment provides $C_{D,p}$.

\subsection{Analysis of the variations of the upstream flow}\label{Analysis of the variations of the upstream flow}

In order to characterize the flow in the settling chamber, measurements of $\meant{u}_x$ and $u_x'$ are carried out using an ensemble of 6 single hot-wires distributed vertically along a profiled bar spanning the settling chamber vertically. This part of the wind tunnel is the largest, with a section of $H\times W=17.64m^2$. The vertical positions of the 6 hot-wires are at $z_\text{hw}/H=\{-0.28, -0.19, -0.04, 0.05, 0.19, 0.29 \}$, see the sketch in~\fref{fig:problem}. The bar can be attached at different spanwise position $y_\text{hw}/H = \{-0.47,-0.35,-0.24,-0.14,-0.03,0.04,0.25,0.36,0.47\}$ so as to probe a large part of the section in a discrete manner. The hot-wires are connected to a constant temperature anenometer and have been calibrated in a preliminary step using a micro jet apparatus. 

\begin{figure}[H]
\centering 
 \begin{subfigure}[b]{0.5\linewidth}
  \centering\includegraphics[width=0.98\linewidth]{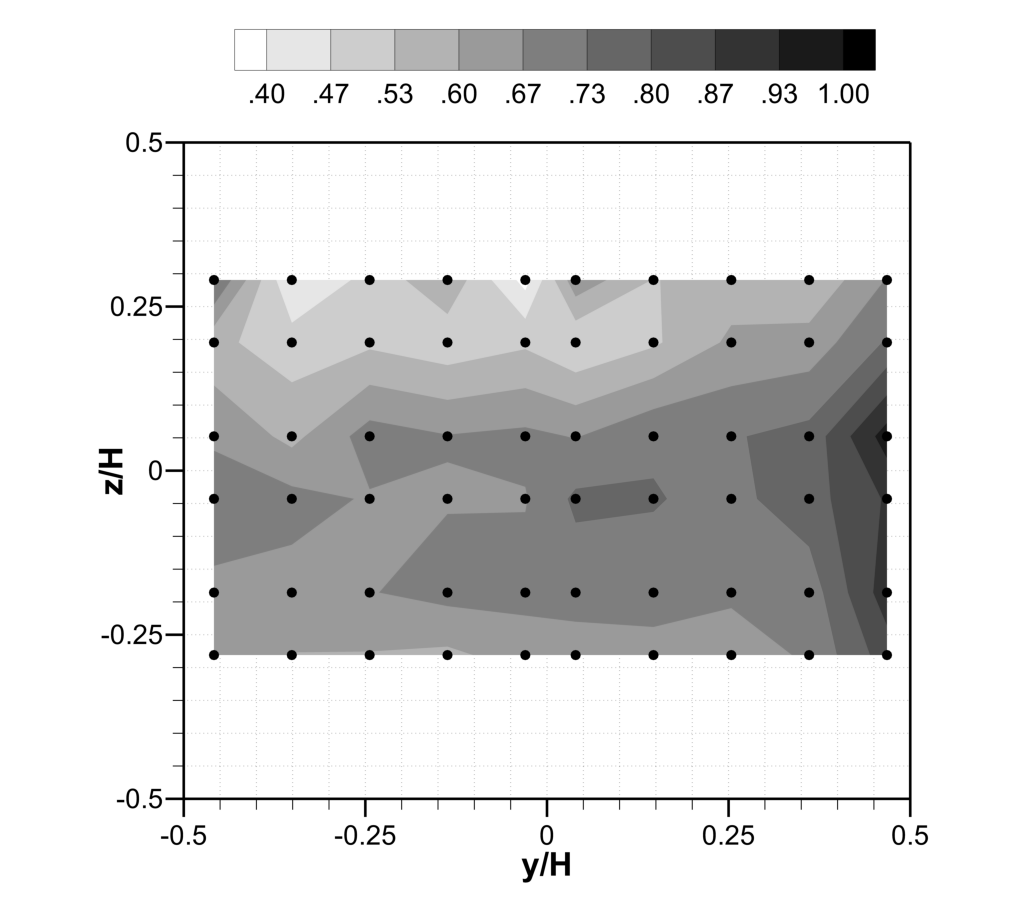} 
\caption{}\label{fig7:a} 
\vspace{2ex}
\end{subfigure}
\begin{subfigure}[b]{0.5\linewidth}
 \centering\includegraphics[width=0.98\linewidth]{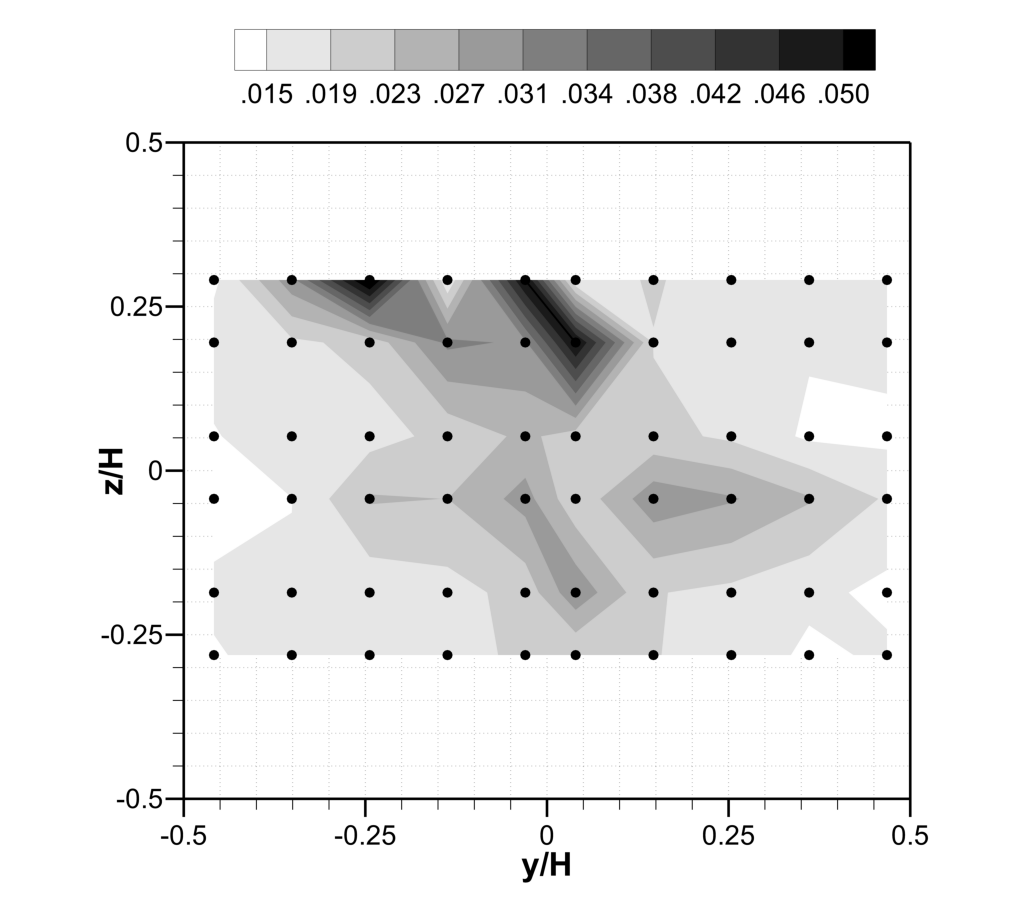}
\caption{}\label{fig7:b} 
    \vspace{2ex}
  \end{subfigure} 
   \caption{Statistical moments of the velocity field in the settling chamber for Mach number \textmd{$\Machts=0.8$} in the test section, obtained from the hot-wire measurements. (a) Time average velocity field $\mu$ normalized upon the typical velocity in the settling chamber \textmd{$u_\its/\RC$}. (b) Turbulence rate \nicola{$CV$}. Bilinear interpolation is used between the scatter points indicated with black dots.  }\label{Statistical moments for M=0.8}
    \end{figure}

Looking at the hot-wire anemometer data as uncertain inputs, the database is processed by evaluating their statistical moments. The average $\mu$, the variance $\sigma^{2}$, the skewness, and the kurtosis of the velocity $u_x(t)$ are thus computed for each hot-wire, over the ensemble of transverse position of the hot-wire mast. Note that in the following the average $\meant{u}_x$ is replaced by $\mu$ ($\mu=\meant{u}_x$) to comply with the regular usage of the statistical framework; \alert{see} \Fref{Statistical moments for M=0.8}. The coordinates are normalized upon the settling chamber height $H$ and velocity upon the typical velocity scale in the settling chamber $u_\its/\RC$. The plot shows a significant velocity inhomogeneity in the section. In particular there is a larger flow velocity in the central part and at the right side while the upper and lower regions yield lower velocities. The spatio-temporally averaged velocity is equal to $0.65 u_\its/\RC$ and maximum difference from this mean is about $0.31 u_\its/\RC$. The turbulence rate $\nicola{\CV}=\sigma / \mu$ in the settling chamber is equal to $2$\% on average with a maximum equal to $5.5$\%.

\subsection{Numerical Simulations}

A RANS model is used to carry out the propagation of this inflow irregular data. The numerical domain is defined in accordance with the experimental one sketched in~\fref{fig:problem}. In particular the account of the settling chamber is decisive to account for the hot-wire data. Indeed the integration of the inlet flow non-homogeneity is made through the inflow boundary condition in the settling chamber. The computational domain is a reduction of the three-dimensional experimental configuration to a two-dimensional setting, as motivated by the large aspect ratio of the cylinder \alert{noticed} before. The principle of the induced modifications of the nozzle shape is detailed later on. 

The simulations are performed using the finite volume CFD solver \textit{elsA}~\cite{cambier2013onera}, considering the Navier-Stokes equations in compressible framework after \nicola{Favre} averaging \nicola{and the relative} decomposition $\velocity=\favre{\velocity} + \velocity''$~\cite{blazek2015}, \alert{which read}
\begin{equation}
    \begin{array}{l}
    \displaystyle{\frac{\partial \meant{\rho}}{\partial t} + \mathrm{\bf Div}(\meant{\rho}\favre{\velocity})} = 0\,, \\
    \displaystyle{\frac{\partial(\meant{\rho}\favre{\velocity})}{\partial t}} + \mathrm{\bf Div}(\meant{\rho}\favre{\velocity}\otimes\favre{\velocity}) = -\boldsymbol{\nabla}\meant{p}+\mathrm{\bf Div}(\meant{\boldsymbol{\tau}}-\meant{\rho\velocity''\otimes\velocity''})\,, \\
    \displaystyle\frac{\partial}{\partial t}\left[\meant{\rho}\left(\favre{E}+\demi\favre{\nord{\velocity''}^2}\right)\right] +\mathrm{\bf Div}\left[\meant{\rho}\left(\favre{E}+\demi\favre{\nord{\velocity''}^2}\right)\favre{\velocity}\right]=-\mathrm{\bf Div}(\meant{p}\favre{\velocity})+\mathrm{\bf Div}\left[(\meant{\boldsymbol\tau}-\meant{\rho\velocity''\otimes\velocity''})\favre{\velocity}\right]-\mathrm{\bf Div}(\meant{\boldsymbol q}+\meant{\rho\enthalpy''\velocity''})\,.
    \end{array}
    \label{eq:rans}
\end{equation}
\alert{Here $\rho$ stands for the flow density, $\boldsymbol{\tau}$ is the viscous stress tensor, $E$ is the total energy, $\enthalpy$ is the specific enthalpy, and $\boldsymbol{q}$ is the heat flux.} The Spalart-Allmaras turbulence model~\cite{spalart2000strategies} is used to close the equations. 

The validation of the numerical simulation against the experimental data at $\Mach_\its=0.8$ is performed using an unsteady RANS simulation (URANS). However because of the cost of such an unsteady simulation and the requirement of a large number of simulations for UQ, only RANS simulations are performed to propagate the inflow uncertainty after the validation phase. 

The \alert{system} of~\nicola{\eref{eq:rans}} is solved using a cell-centered finite volume spatial discretization on structured multiblock meshes. All the simulations are carried out using a multigrid approach and the spatial scheme proposed by Jameson \emph{et al.} \cite{jameson1981numerical} is used for the conservative variables. The second-order dissipation coefficient $\chi_2$ and the fourth-order linear dissipation coefficient $\chi_4$ are set to $0.5$ and  $0.016$, respectively. For the implicit stage, a lower/upper symmetric successive over-relaxation (LU-SSOR) numerical scheme \cite{yoon1986lu} is associated with an Euler backward time-integration scheme, ensuring fast convergence rates. For the turbulent variables, a first order version of the Roe numerical scheme is used with a Harten entropic correction coefficient set to $0.01$ and the minmod limiter. The transition of the boundary layer at the surface of the cylinder is let free with the location computed using a supersonic extension of the Arnal-Habiballah-Delcourt (AHD) \cite{arnal1989transition} criterion combined with the Gleyzes \emph{et al.} \cite{gleyzes1985theoretical} criterion. The flow turns turbulent whenever one of these two criteria activates. The AHD criterion is determined by the N factor of the freestream which is set to 5.5~\cite{brion2020laminar}.

The numerical model is two-dimensional and accounts for the flow in the ensemble made of the settling chamber, nozzle, and test section, as shown in~\fref{fig:problem}. The test section being rectangular the two-dimensional numerical setup results from neglecting the effect of viscosity at the lateral walls, hence replacing them by periodic boundary conditions. The nozzle however can not be reduced to a two-dimensional setting since its shape evolves three-dimensionally from the settling chamber to the test section. Hence the two-dimensional numerical setup \alert{considers} a planar nozzle of width equal to the test section width ($w=0.804m$) instead of the real one. The same height of settling chamber and same nozzle geometry for the upper and lower surfaces are taken in the modified geometry. The change in nozzle geometry results in a change of velocity magnitude in the settling chamber with a ratio equal to $\RD=4.975$ between the two-dimensional configuration and the real three-dimensional settling chamber. A velocity difference remains all the way down to the test section where experiments and numerics resume to agreement. 

\begin{figure}[H]
\minipage{0.5\textwidth}
\centering\includegraphics[width=0.9\linewidth]{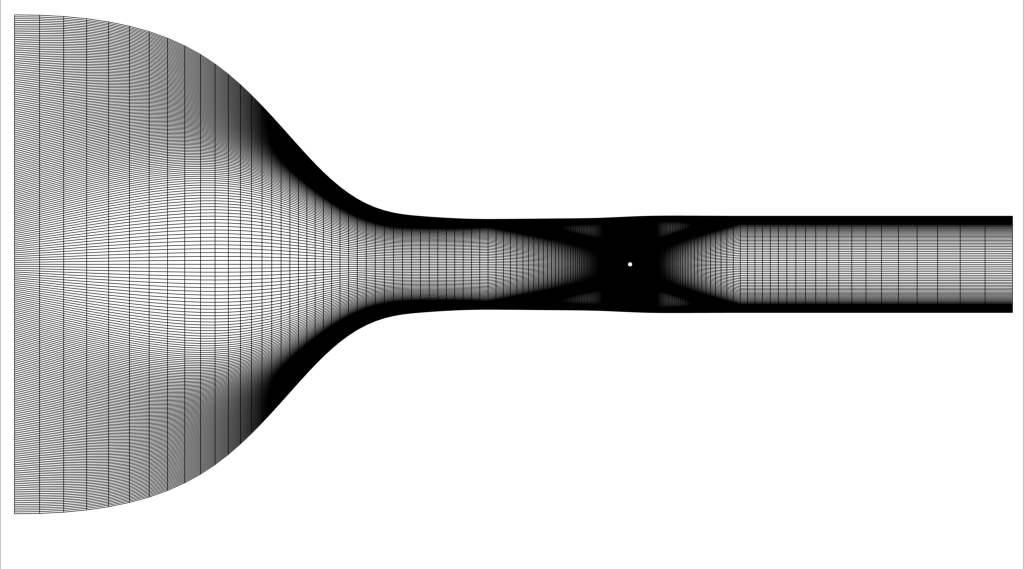}
  \subcaption{}
 \endminipage\hfill
\minipage{0.5\textwidth}
\centering\includegraphics[width=0.9\linewidth]{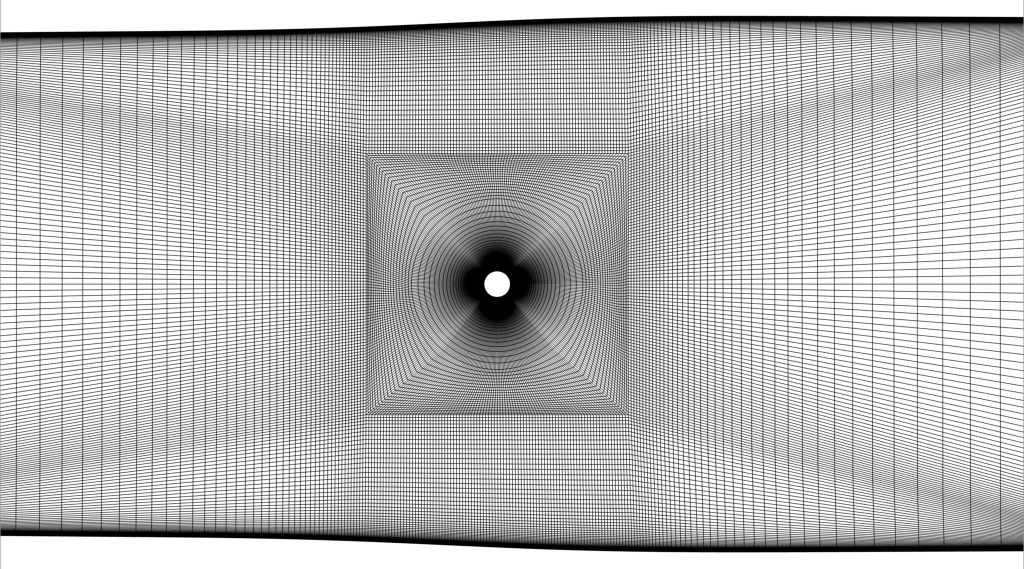}
  \subcaption{}
 \endminipage\hfill
 \caption{View of the two-dimensional computational grid used for the RANS simulations. (a) Entire mesh from settling chamber to test section. (b) Closed-up view of the grid refinement generated around the cylinder. Note the adapted walls above and below the cylinder, which expand symmetrically in reaction to the presence of the cylinder so as to maintain a constant Mach number. }
\label{fig:grid}
\end{figure}

\Fref{fig:grid} shows the grid for the flow domain and a closed up view around the cylinder in the test section with deformed upper and lower walls. The shape of the adaptive walls are integrated into the mesh by mesh deformation of the initially rectilinear domain.

\begin{figure}[H]
\centering\includegraphics[width=0.54\textwidth]{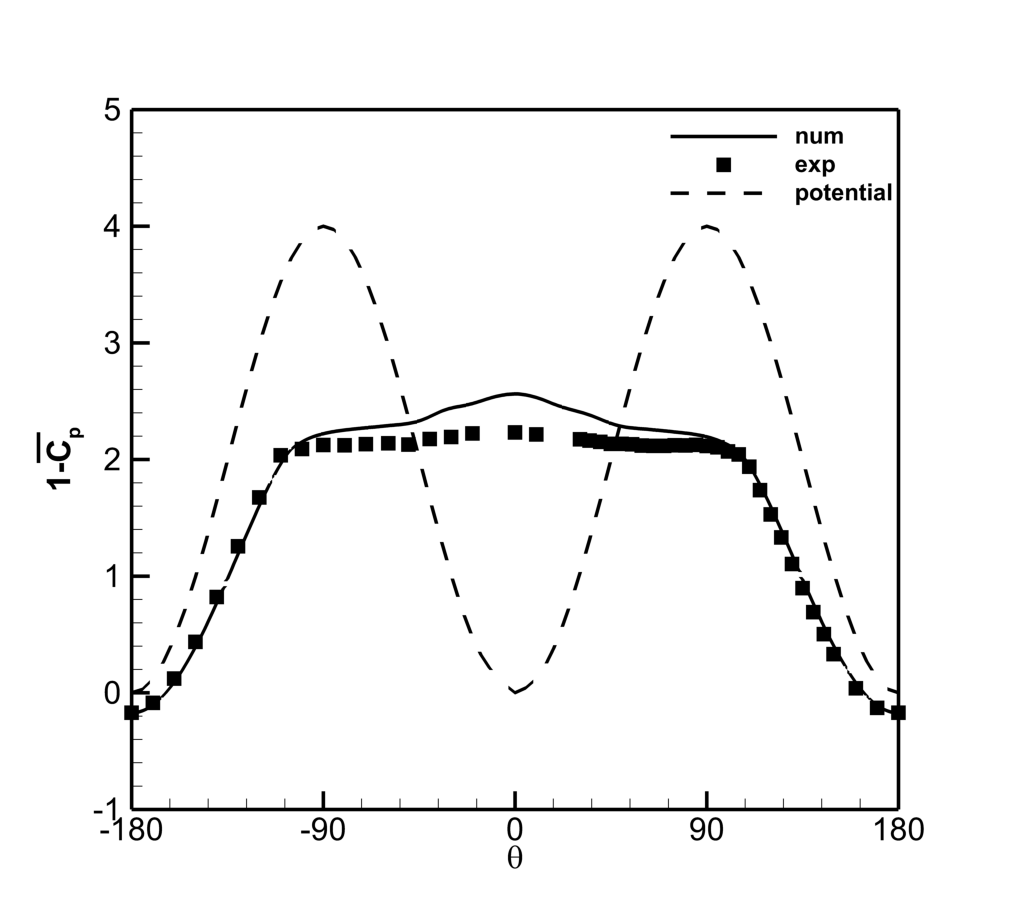} 
\caption{Comparison of the pressure distributions around the cylinder at \textmd{$\Machts=0.8$} between experiment, simulation and potential flow solution $1-\meant{C}_p=4\sin^2\theta$.}
\label{fig:cpcomparison}
\end{figure}

Inflow conditions take into account the stagnation conditions $P_\iin$ and $H_\iin=c_p T_\iin$ as provided from the experiment and the flow is installed by setting a back pressure at the outflow boundary. This back pressure is adjusted manually to match the pressure distribution at the cylinder surface. The comparison between the experimental pressure distribution and the numerical prediction is shown in~\fref{fig:cpcomparison}, along with the potential flow, incompressible solution, as reference. With the adjusted back pressure, the experimental pressure distributions are well matched overall, with a close to perfect agreement at the front part of the cylinder and slight discrepancy after flow separation, certainly as a consequence of the strong sensitivity of the rear pressure level to the precise location of the separation. The potential flow solution indicates some evidence of the viscous effects causing milder pressure variations in the front part due to boundary layer effects and flow separation at the back. The time-averaged value of the drag coefficient obtained from the URANS simulation is $\meant{C}_D=1.84$ and the experimental pressure contribution is $\meant{C}_{D,p}=1.26$. 

\subsection{Numerical results and comparisons against experiment}\label{Numerical results}

\begin{figure}[H]
  \begin{subfigure}[b]{0.5\linewidth}
    \centering\includegraphics[width=0.9\linewidth]{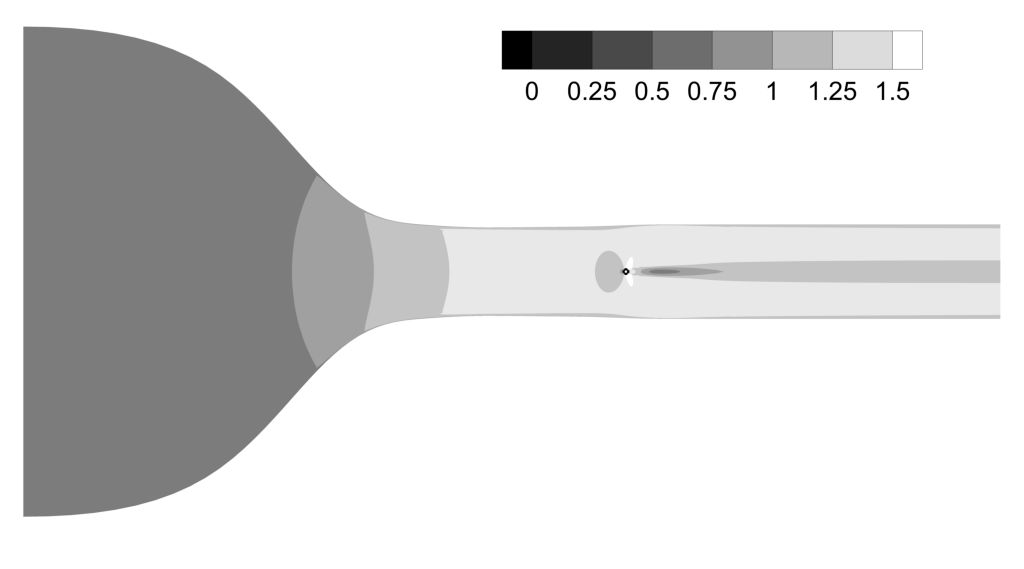}
    \caption{} 
        \vspace{2ex}
  \end{subfigure} 
  \begin{subfigure}[b]{0.5\linewidth}
    \centering\includegraphics[width=0.9\linewidth]{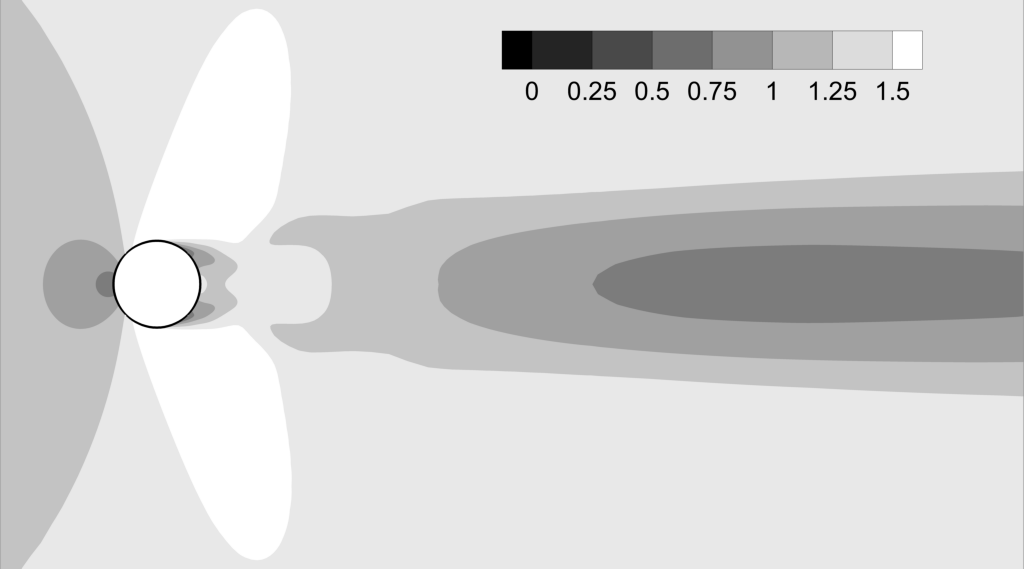}
    \caption{} 
        \vspace{2ex}
  \end{subfigure}
    \caption{Time-averaged flow field showing iso-contours of the Mach number for \textmd{$\Machts=0.8$} obtained from the URANS simulation. (a) View of the entire computational domain. (b) Closed-up view in the \alert{vicinity} of the cylinder showing the wake pattern.}
    \label{contours}
\end{figure}

\begin{figure}[H]
\minipage{0.5\textwidth}
  \includegraphics[width=0.98\linewidth]{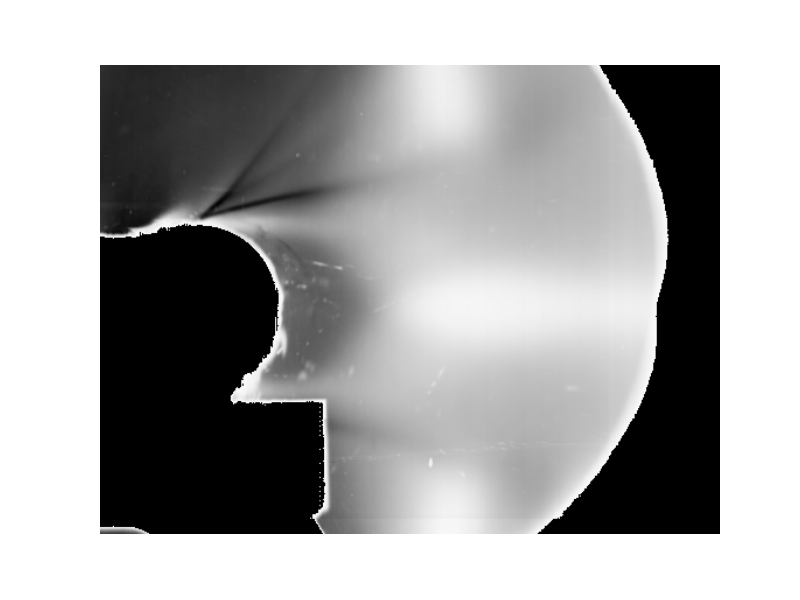}
   \subcaption{} 
\endminipage\hfill
\minipage{0.5\textwidth}
  \includegraphics[width=0.98\linewidth]{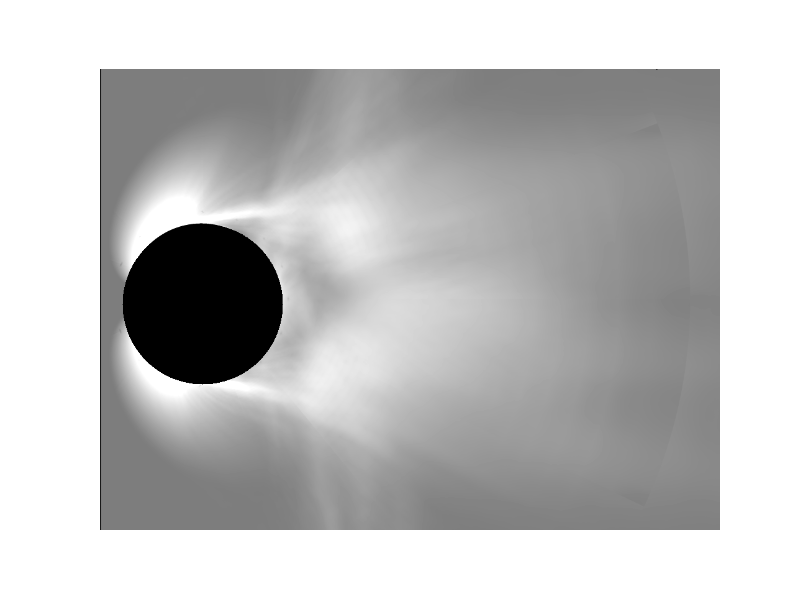}
  \subcaption{} 
\endminipage
 \caption{Time-averaged Schlieren visualisations of the flow past the cylinder at \textmd{$\Machts=0.8$}. (a) Experimental vs. (b) numerical results obtained from averaging of the URANS simulation.}
 \label{grad_rho}
\end{figure}

The time-averaged Mach number field from the URANS simulations \alert{is} illustrated in~\fref{contours}. The flow past the cylinder is accelerated at the front part, reaching supersonic speeds in the region of the cylinder apex above and below. The supersonic zone is terminated by a straight shock wave. The flow separates shortly after the cylinder apex, generating an important wake that grows in width downstream. Another view of the time-averaged flow field is provided in~\fref{grad_rho} using Schlieren visualisations accounting for the gradient of flow density. The numerical Schlieren are readily compared to the experimental ones, showing similar representation of the flow, notably the shock wave, the separated boundary layer and the recirculation area behind the cylinder.
\vspace{5mm}

\section{Uncertainty quantification}\label{sec:UQ}

The aim of this section is to characterize the influence of the uncertainties that affect the inflow wind tunnel velocity on the transonic flow around \alert{the} test cylinder and, in particular, on some aerodynamic quantities of interest, here its drag coefficient. In order to do this, a surrogate model is an efficient solution, and it is frequently used in CFD to perform optimization and uncertainty quantification (UQ); \bibrev{see \emph{e.g.} \cite{DOD09} and references therein}.

The principle of a surrogate model relies on an interpolation or regression procedure to estimate a scalar or a vector field, using a sampling dataset made up of the outputs of some complex process. Generally, this process can be extremely expensive to run and the surrogate model allows us to emulate it, obtaining rapidly output samples without any extra computational costs. In this research a polynomial chaos expansion (PCE) \bibrev{\cite{ghanem2003stochastic,soize2004physical,wiener1938homogeneous,xiu2002wiener,le2010spectral}} is applied to the foregoing wind tunnel inflow problem. The PCE is a powerful tool for constructing a spectral-like surrogate model of a complex process $f$ (e.g. a CFD computation) depending on $D$ random input parameters $\Xiv$. It consists in expanding the output quantity of interest (QoI) of that process onto a basis of orthogonal polynomials. Namely, if $\QoI=f(\Xiv)$ is a QoI depending on the real-valued random inputs $\Xiv$ characterized by their probability density function (PDF) \nicola{$\xiv\to\pdf(\xiv)$} on $\Rset^D$, its polynomial chaos expansion reads:
\begin{equation}\label{eq:PCE}
\QoI = f(\Xiv) \simeq \sum^{\infty}_{j=0} c_j \Psi_j(\Xiv)\,,
\end{equation} 
where $\Psi_j$ is an element of a family of orthogonal polynomials with respect to \nicola{$\pdf(\xiv)$}, and $c_j$ is the associated expansion coefficient for that polynomial. Practically, obtaining these coefficients constitutes most of the work required to implement the method, considering that as many other surrogate models PCE suffers from the so-called "curse of dimensionality": the computational costs increases exponentially with the number $D$ of uncertain input parameters. For this reason, when considering a large space of uncertain parameters, efficient algorithms are needed to obtain an accurate surrogate representation of the parametric output $\QoI$. 

Two approaches for computing the coefficients $c_j$ of the PCE of \eref{eq:PCE} are typically considered: (i) a projection approach by which they are computed by structured (Gauss) quadratures; and (ii) a regression approach by which they are computed by minimizing some error tolerance.

\subsection{Definition of the uncertainties}\label{Definition of the uncertainties}

Since numerical simulations have been performed in two dimensions, the inflow data for $\Machts = 0.8$ and for one position $y_\text{hw}/H=-0.351$ spanwise are chosen. The uncertain parameters in the present research are the six velocity fluctuations measured at the inflow wind tunnel section; see \sref{Analysis of the variations of the upstream flow}. Computing the second-order statistical moments of each random variable in \tref{inflow velocity uncertainties} and their histogram, they seem to follow a Gaussian distribution.

\begin{table}[H]
\begin{center}
\begin{tabular}{|c|c|c|c|c|c|c|}
\hline
$-$		&	$U_{1}$	& $U_{2}$ & $U_{3}$ & $U_{4}$ & $U_{5}$ & $U_{6}$\\
\hline
$\mu \;\; \RC/u_\its$	&	0.60		&	0.64 & 0.69 & 0.58 & 0.50 & 0.40 \\
\hline
$\sigma$	& 0.10	&	0.10 & 0.10 & 0.10 & 0.09 & 0.11\\
\hline
\nicola{$\CV$} (\%)	&	1.7	&	1.5 & 1.5 & 1.8 & 1.8 & 2.9 \\
\hline
\end{tabular}
\end{center}
\caption{Inflow velocity fluctuations and their statistical moments: mean $\mu$ normalized on the typical scale of velocity in the settling chamber \textmd{$u_\its/\RC$}, standard deviation $\sigma$, and coefficient of variation $\CV=\sigma/\mu$ for the position $y_\mathrm{hw}=-0.351$. }\label{inflow velocity uncertainties}
\end{table}

In order to check this hypothesis, the Kullback-Leibler (KL) divergence, or relative entropy can be used. It is a measure of how a probability distribution is different from another reference probability distribution \bibrev{\cite{kullback1951information}}. For the distributions ${\mathcal P}$ and ${\mathcal Q}$ of a continuous random variable defined on the same probability space, the KL divergence is: 
\begin{equation}\label{eq:KLdiv}
D_\text{KL}({\mathcal P}\mid\mid {\mathcal Q}) = \int_{-\infty}^{\infty}\mathrm{p}(\xi)\log \left(\dfrac{\mathrm{p}(\xi)}{\mathrm{q}(\xi)}\right)d\xi
\end{equation}
where $\mathrm{p}$ and $\mathrm{q}$ denote the PDFs of ${\mathcal P}$ and ${\mathcal Q}$. In other words, it is the expectation of the logarithmic difference between the PDFs $\mathrm{p}$ and $\mathrm{q}$, where the expectation is taken using the PDF $\mathrm{p}$. Using the KL divergence, it is possible to quantify the distance between the distributions of the inflow velocity fluctuations and Gaussian distributions $\NLaw(\mu,\sigma)$ with the same mean $\mu$ and standard deviation $\sigma$. \tref{KL divergence} shows that Gaussian distributions can reasonably be associated to these velocity fluctuations. Here the Gaussian PDF is $\npdf(\xi;\mu,\sigma)=\frac{1}{\sigma\sqrt{2\pi}}\exp[-\demi(\frac{\xi-\mu}{\sigma})^2]$, where the mean $\mu$ and standard deviation $\sigma$ are given in \tref{inflow velocity uncertainties} for the six inflow velocity random fluctuations.

\begin{table}[H]
\begin{center}
\begin{tabular}{|c|c|c|c|c|c|c|}
\hline
$-$		& $U_{1}$ & $U_{2}$ & $U_{3}$ & $U_{4}$ & $U_{5}$ & $U_{6}$\\
\hline
$D_\text{KL}(U_i\mid\mid \NLaw)$ [$\times$ 1e-4]	&	3 & 2	 & 2 & 3 & 3 & 8 \\
\hline 
\end{tabular}
\end{center}
\caption{KL divergence with a Gaussian distribution computed for the six random variables $U_1,U_2,\dots U_6$.}\label{KL divergence}
\end{table}

\subsection{Polynomial chaos surrogate model}\label{Chaos Polynomials surrogate model}

Polynomial chaos surrogate models are considered for QoIs $\QoI$, here the drag of the cylinder. The random input parameters $\Xiv=(U_1,U_2,\dots U_6)$ form a random vector of $\Rset^6$ (the parameter space dimension is thus $D=6$) with independent coordinates. Hence its PDF is the product of the PDFs of each individual coordinate, \nicola{$\pdf(\xiv)=\prod_{i=1}^6\pdf_i(\xi_i)$}, where the PDF \nicola{$\pdf_i$} of the $i$-th velocity fluctuation measured at the inflow wind tunnel section is a Gaussian density \nicola{$\pdf_i(\xi)=\npdf(\xi;\mu_i,\sigma_i)$} with mean $\mu_i$ and standard deviation $\sigma_i$ given in \tref{inflow velocity uncertainties} for $i=1,2,\dots 6$. The family of orthogonal polynomials with respect to \nicola{$\pdf(\xiv)$} is constituted by the multi-dimensional polynomials which are the products of the one-dimensional polynomials in each coordinate $U_i$ orthogonal with respect to \nicola{$\pdf_i(\xi)$}. That is, one has:
\begin{equation}\label{tensor product}
\Psi_{\bf j}(\Xiv)=\prod_{i=1}^{D}\psi_{j_{i}}(U_i)\,,
\end{equation}
where ${\bf j}=(j_1,j_2,\dots j_D)$ is actually a multi-index in $\Nset^D_0=\Nset^D\cup\{{\bf 0}\}$, and the one-dimensional polynomials $\psi_j$ satisfy:
\begin{equation}\label{eq:1DorthoPol}
\inner{\psi_j,\psi_k}:=\int_\Rset\psi_j(\xi)\psi_k(\xi)\nicola{\pdf_i(\xi)}d\xi=\esp[\psi_j(U_i)\psi_k(U_i)]=\delta_{jk}\,.
\end{equation}
Here $\esp[\cdot]$ stands for mathematical expectation (mean), and $\delta_{jk}=1$ if $j=k$ and $\delta_{jk}=0$ otherwise stands for the Kronecker symbol. The one-dimensional polynomials $\psi_k$ are actually orthonormal with respect to \nicola{$\pdf_i(\xi)$} with the definition above, and the multi-dimensional polynomials $\Psi_{\bf j}$ are consequently orthonormal with respect to \nicola{$\pdf(\xiv)$}:
\begin{equation}\label{eq:DDorthoPol}
\inner{\Psi_{\bf j},\Psi_{\bf k}} = \int_{\Rset^D}\Psi_{\bf j}(\xiv)\Psi_{\bf k}(\xiv)\nicola{\pdf(\xiv)}d\xiv = \esp\left[\Psi_{\bf j}(\Xiv)\Psi_{\bf k}(\Xiv)\right]  =\delta_{{\bf j}{\bf k}}\,,
\end{equation}
where $\delta_{{\bf j}{\bf k}}=\delta_{j_1k_1}\delta_{j_2k_2}\dots\delta_{j_Dk_D}$. The PCE of \eref{eq:PCE}:
\begin{equation}\label{PC_norm}
\QoI=f(\Xiv)\simeq\sum_{{\bf j}\in \Nset_0^D}c_{\bf j}\Psi_{\bf j}(\Xiv)
\end{equation}
contains infinitely many terms, and for the purpose of numerical computation the summation should be truncated. Introducing the total order \nicola{$\torder$} of the multi-variate polynomials such that $\norm{{\bf j}}_1=\sum_{i=1}^Dj_i\leq \nicola{\torder}$, the number of terms in the expansion (\ref{PC_norm}) is:
\begin{equation}\label{eq:Pp1}
\porder+1=\binom{\nicola{\torder}+D}{D}=\dfrac{(\nicola{\torder}+D)!}{\nicola{\torder}!D!}
\end{equation}
and \eref{PC_norm} for $\norm{{\bf j}}_1\leq\torder$ reads:
\begin{equation}\label{PC_norm1}
\QoI\simeq g_\porder(\Xiv):=\sum_{j=0}^\porder c_j\Psi_j(\Xiv)
\end{equation}
re-indexing the multi-variate polynomials of total order less than \nicola{$\torder$} with a single index $j=0,1,\dots\porder$.

\subsection{Orthonormal polynomial basis}\label{Polynomials orthogonal basis}

In the previous section the one-dimensional orthogonal polynomials have been introduced through \eref{eq:1DorthoPol}. Starting from Wiener work about Gaussian random variables \cite{wiener1938homogeneous}, the Askey scheme \cite{askey1985some} is invoked in \cite{xiu2002wiener} to extend polynomial chaos families to different processes in order to apply this approach whatever the distribution of the uncertain parameter is; see also \cite{soize2004physical}. Therefore, in relation to the PDFs of the random input parameters, a particular family is chosen. Here the six independent random inputs have Gaussian distributions and thus, Hermite polynomials will be used to construct the \alert{polynomial surrogate model}. In this study, the probabilists Hermite polynomials $H_j$ are considered, using the following general representation:
\begin{equation}
H_j(\xi)= \frac{(-1)^j}{\npdf_0(\xi)}\frac{d^j\npdf_0(\xi)}{d\xi^j}
\end{equation}
where $\npdf_0(\xi):=\npdf(\xi;0,1)=\frac{1}{\sqrt{2\pi}}\iexp^{\frac{-\xi^2}{2}}$ and the $j$-th order Hermite polynomial is a polynomial of degree $j$. These polynomials are orthogonal with respect to the normal density $\npdf_0$, that is:
\begin{equation}\label{orthogonality}
\int_\Rset H_j(\xi)H_k(\xi)\npdf_0(\xi)d\xi = j!\delta_{jk}\,,
\end{equation}
such that in their normalized version $\psi_j(\xi):=(j!)^{-\demi}H_j(\xi)$ the orthonormality relationship (\ref{eq:1DorthoPol}) is fulfilled. In the following sections, the different strategies used to compute the coefficients of the series are outlined.

\subsection{Projection approach}\label{Projection approach}

We now turn to the computation of the expansion coefficients $c_j$ in \eref{PC_norm1}, considering at first the projection approach. It is used to compute a reference solution in order to validate the results obtained with compressed sensing, which will be exposed in \sref{Regression approach}. As seen in \sref{Chaos Polynomials surrogate model}, the output QoI $\QoI$ being represented by the PCE (\ref{PC_norm1}), computing the inner product (\ref{eq:DDorthoPol}) yields $\inner{g_\porder,\Psi_k} = \sum^\porder_{j=0} c_j\inner{\Psi_j,\Psi_k}=\sum^\porder_{j=0} c_j\delta_{jk}$, hence:
\begin{equation}\label{projection}
c_j = \int_{\Rset^D} g_\porder(\xiv)\Psi_j(\xiv)\nicola{\pdf(\xiv)}d\xiv\,;
\end{equation}
that is, it consists in a projection of the QoI $\QoI\simeq g_\porder(\Xiv)$ onto the polynomial basis. At this stage, one can remark from this result that the expansion coefficients are related to the second-order statistical moments of the QoI. Indeed its average $\mu_\porder$ can be computed as:
\begin{equation}\label{average_coeff}
\mu_\porder=\esp[g_\porder]=\inner{g_\porder,1}=c_0
\end{equation}
since $\Psi_0(\xiv)=1$, and its mean-square root $\sigma_\porder^2$ as: 
\begin{equation}\label{sd_coeff}
\begin{split}
\sigma_\porder^2 &=\esp[g_\porder^2]-\mu_\porder^2 \\
&=\sum_{j=0}^\porder\sum_{k=0}^\porder c_jc_k\inner{\Psi_j,\Psi_k}-c_0^2 \\
&=\sum_{j=1}^\porder c_j^2\,.
\end{split}
\end{equation}

In order to compute the integral (\ref{projection}), we use a Gauss quadrature (GQ) rule that is adapted to the condition of orthogonality (\ref{orthogonality}) in one dimension. Since we can always fit a $Q-1$ degree polynomial to a set of $Q$ points, the following integral can be evaluated exactly:
\begin{equation}\label{GQ}
\int_\Rset h(\xi)\npdf_0(\xi)d\xi = \sum_{i=1}^Q w_i h(\xi_i)
\end{equation}
by carefully choosing the weights and abscissas $(w_i,\xi_i)_{1\leq i\leq Q}$, provided that the function $\xi\to h(\xi)$ defined on $\Rset$ is a polynomial of degree not greater than $2Q-1$. $(w_i,\xi_i)_{1\leq i\leq Q}$ are the $Q$ Gauss-Hermite quadrature weights and points \cite{gubner2009gaussian} associated with the weight $\npdf_0$ defined on $\Rset$ (a PDF in the present case). A Gauss-Hermite quadrature rule in $D$ dimensions can subsequently be constructed by full tensorization of the one-dimensional rule above, yielding:
\begin{equation}\label{cjGQ}
c_j= \int_{\Rset^D}g_\porder(\xiv)\Psi_j(\xiv)\nicola{\pdf(\xiv)}d\xiv\simeq\sum_{i=1}^Q w_i g_\porder(\xiv_i)\Psi_j(\xiv_i)=c_j^Q
\end{equation}
where $(w_i,\xiv_i)_{1\leq i\leq Q}$ are the $Q=\prod_{d=1}^D Q_d$ Gauss-Hermite nodes and weights in $D$ dimensions when $Q_d$ nodes are considered for the $d$-th dimension, and for $1\leq i_d\leq Q_d$:
\begin{equation}
w_i=\prod_{d=1}^D w_{i_d}\,,\quad\xiv_i=(\xi_{i_1},\xi_{i_2},\dots\xi_{i_D})\,.
\end{equation}
In particular, the present problem involves $D=6$ random variables and we use $Q_d=4$ Gauss-Hermite points for each random dimension. We are thus able to integrate exactly the orthogonality rule (\ref{eq:DDorthoPol}) in $D=6$ dimensions for polynomials up to a total degree $\torder=3$.

\subsection{Regression approach}\label{Regression approach}

The regression approach adopted in this work is based on the general idea of reconstructing a generic signal taking into account only few evaluations of it. Indeed, many natural signals have concise representations when expressed in a convenient basis. For this reason, they can be considered as sparse or compressible in the terminology adopted in the theory of compressed sensing, or compressive sampling (CS) \bibrev{\cite{CAN06,DON06}}. Adapting this idea to the UQ framework outlined above, the starting observation is that many stochastic problems are characterized by a sparse chaos representation. A PCE is considered as sparse if a small but unknown subset of the polynomial basis is able to approximate efficiently the QoI (in a suitable sense). In particular, this is expected to be the case for stochastic processes with a large number of random input variables \cite{todor2007convergence}, where the PCE is supposed to exhibit sparsity in a small fraction of its coefficients. From this point of view, CS represents an efficient route for the reconstruction of sparse PCE solutions, aiming at selecting a few basis polynomials with great impact on the model response \cite{tsilifis2019compressive}.

Using $N$ samples of the random input variables $\Xiv$ generated by a Monte-Carlo method, namely $(\xiv_1,\xiv_2,\dots \xiv_N)$, one value $\QoI_i$ of the output QoI is obtained for each sample $\xiv_i$ (with $i=1,2,\dots N$) by running the CFD solver and, thus, one value of the truncated PCE (\ref{PC_norm1}). In a compact way, gathering all PCEs for all samples the following linear system is formed:
\begin{equation}
\boldsymbol{\QoI}= [ \boldsymbol{\Psi}]\boldsymbol{c}
\end{equation}
where $\boldsymbol{\QoI} =(\QoI_1,\QoI_2,\dots\QoI_N)^\itr$, $\boldsymbol{c} = (c_1,c_2,\dots c_\porder)^\itr$, and $[\boldsymbol{\Psi}]$ is the $N\times \porder$ measurement matrix with $[\boldsymbol{\Psi}]_{ij} = \Psi_j(\xiv_i)$, where typically $N\ll \porder$. The system has to be solved in favour of the vector of the expansion coefficients $\boldsymbol{c}$, but it is an undetermined system of linear equations and, from a mathematical point of view, it would have an infinite number of solutions. However CS theory states that imposing a "constraint of sparsity" whereby only solutions which have a small number of non-zero coefficients are allowed, an unique solution can be recovered with a probability of almost $1$. In order to do this, there exist a wide variety of methods for sparse recovery of signals from a set of incomplete (under-determined) random measurements, for example the $\ell_{1}$-minimization. In particular, considering that the $\kappa^\text{th}$ (total) order polynomial chaos representation $g_\porder$ of the output QoI $\QoI$ is not necessarily complete or exact, a relaxed optimization problem called Basis Pursuit Denoising (BPDN) can be considered \bibrev{\cite{CHE98}}:
\begin{equation}\label{cjCS}
\alert{\boldsymbol{c}^\star} = \underset{\boldsymbol{c}}{\text{argmin}}\nord{\boldsymbol{c}}_1 \quad\text{subject to}\quad\nord{[\boldsymbol{\Psi}]\boldsymbol{c} - \boldsymbol{\QoI}}_{2}\leq\epsilon\,,
\end{equation}
where $\epsilon$ is an $L^2$-error tolerance for the truncated PCE (\ref{PC_norm1}), and $\nord{\boldsymbol{c}}_1=\sum_{j=0} ^\porder\norm{c_j}$.

\section{Application to the wind tunnel experiments}\label{sec:application}

Implementing the theoretical framework outlined in \sref{sec:UQ}, the chaos expansion coefficients \alert{$\boldsymbol{c}^Q$} obtained with the projection approach, and \alert{$\boldsymbol{c}^\star$} obtained using the regression approach, are computed. Since the total order of the multi-variate polynomials has been fixed to $\torder=3$ and the dimension of the parameters set is $D=6$, \alert{$\porder=84$} expansion coefficients have to be computed; see \eref{eq:Pp1}. Since $4$ Gauss-Hermite quadrature points are used for each dimension of the parameters set, the projection approach needs $Q=4^6=4096$ CFD simulations to compute the PCE coefficients by \eref{cjGQ}. In the regression approach, we considered $N=21$ evaluations to compute the PCE coefficients by \eref{cjCS}. For that purpose we use the Spectral Projected Gradient Algorithm (SPGL) developed by van den Berg \& Friedlander \cite{van2009probing} and implemented in the package \texttt{SPGL1} \cite{van2007spgl1} to solve this $\ell_1$-minimization problem.

\subsection{Numerical simulation: Gauss-Hermite points and Monte Carlo sampling}\label{Numerical simulation - Gauss-Hermite points}

Once the normalized Hermite polynomials have been computed at the $Q=4096$ points $(\xiv_1,\xiv_2,\dots\xiv_Q)$ from a Gauss-Hermite quadrature set, or at $N=21$ points $(\xiv_1,\xiv_2,\dots\xiv_N)$ from a random sampling set by the Monte-Carlo method, the coefficients $\boldsymbol{c}^Q=(c_0^Q,c_1^Q,\dots c_\porder^Q)^\itr$ obtained from \eref{cjGQ}, or \alert{$\boldsymbol{c}^\star=(c_0^\star,c_1^\star,\dots c_\porder^\star)^\itr$} obtained from \eref{cjCS}, can be computed provided that the QoI vector $\boldsymbol{\QoI}$ is known. Here the QoI is the drag coefficient $\QoI\equiv C_D(\Xiv)$ for the six velocity random fluctuations $\Xiv\in\Rset^6$ measured at the inflow wind tunnel section (see \tref{inflow velocity uncertainties}). 

\begin{figure}[H] 
\centering
\begin{subfigure}[b]{0.5\linewidth}
\centering\includegraphics[width=0.75\linewidth]{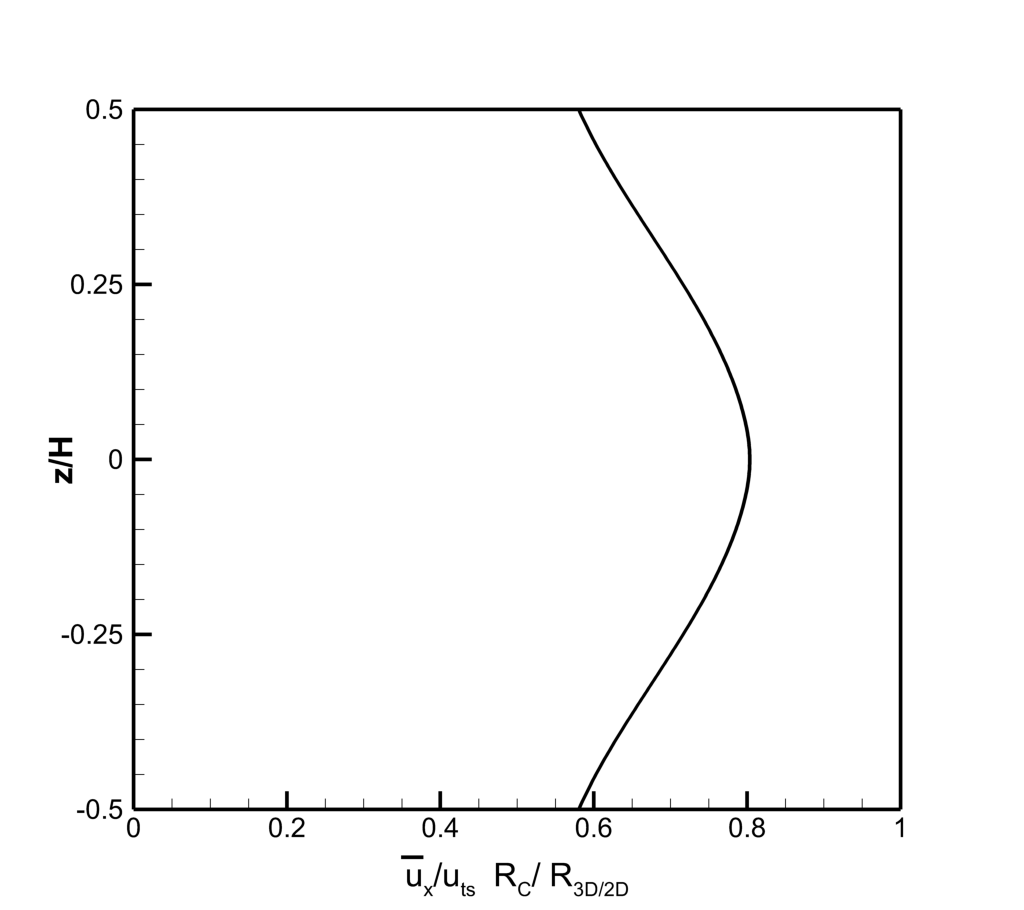}
\caption{}
\vspace{1ex}
\end{subfigure}
\begin{subfigure}[b]{0.5\linewidth}
\centering\includegraphics[width=0.75\linewidth]{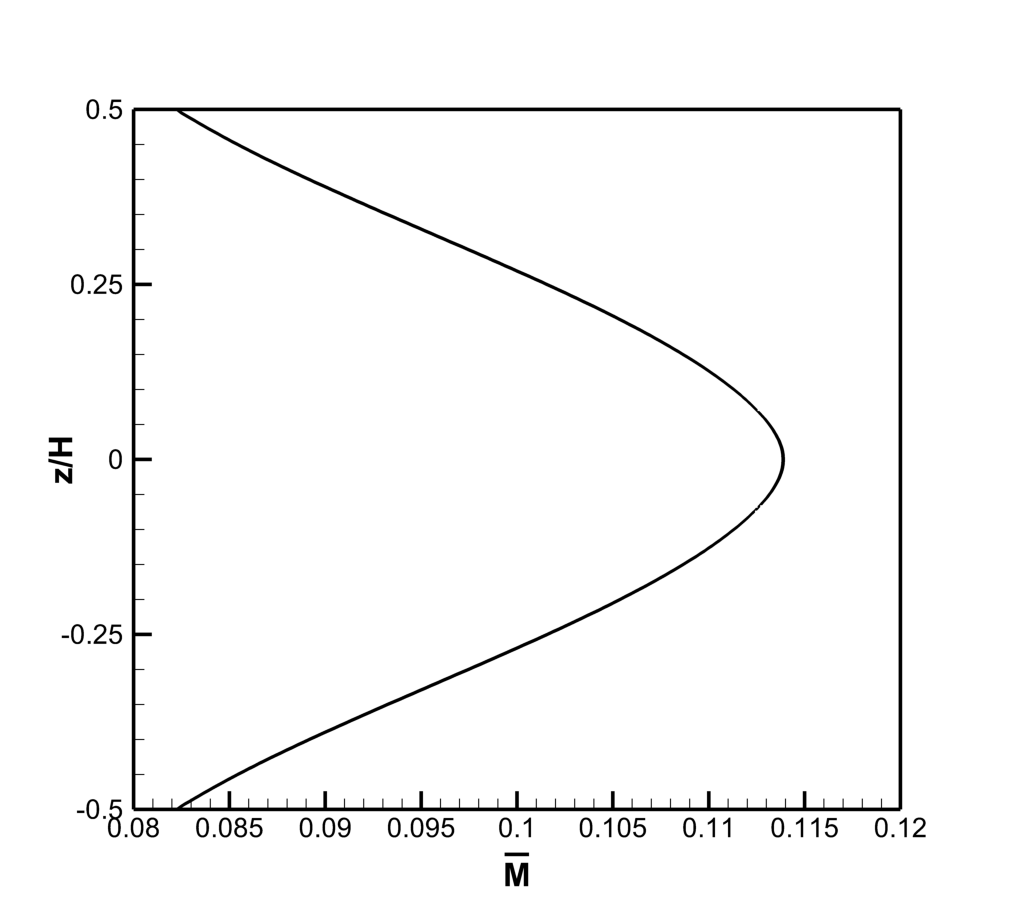}
\caption{}
\vspace{1ex}
\end{subfigure} 
\begin{subfigure}[b]{0.5\linewidth}
\centering\includegraphics[width=0.75\linewidth]{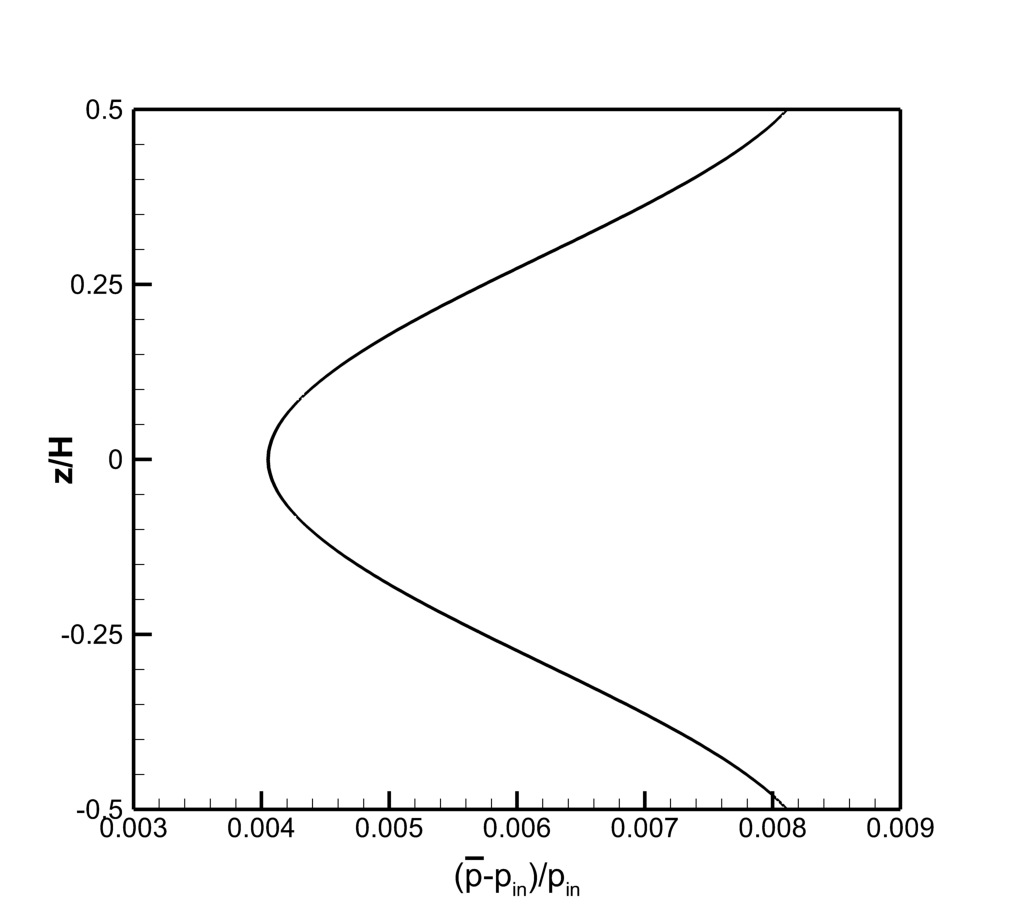}
\caption{}
\end{subfigure}
\begin{subfigure}[b]{0.5\linewidth}
\centering\includegraphics[width=0.75\linewidth]{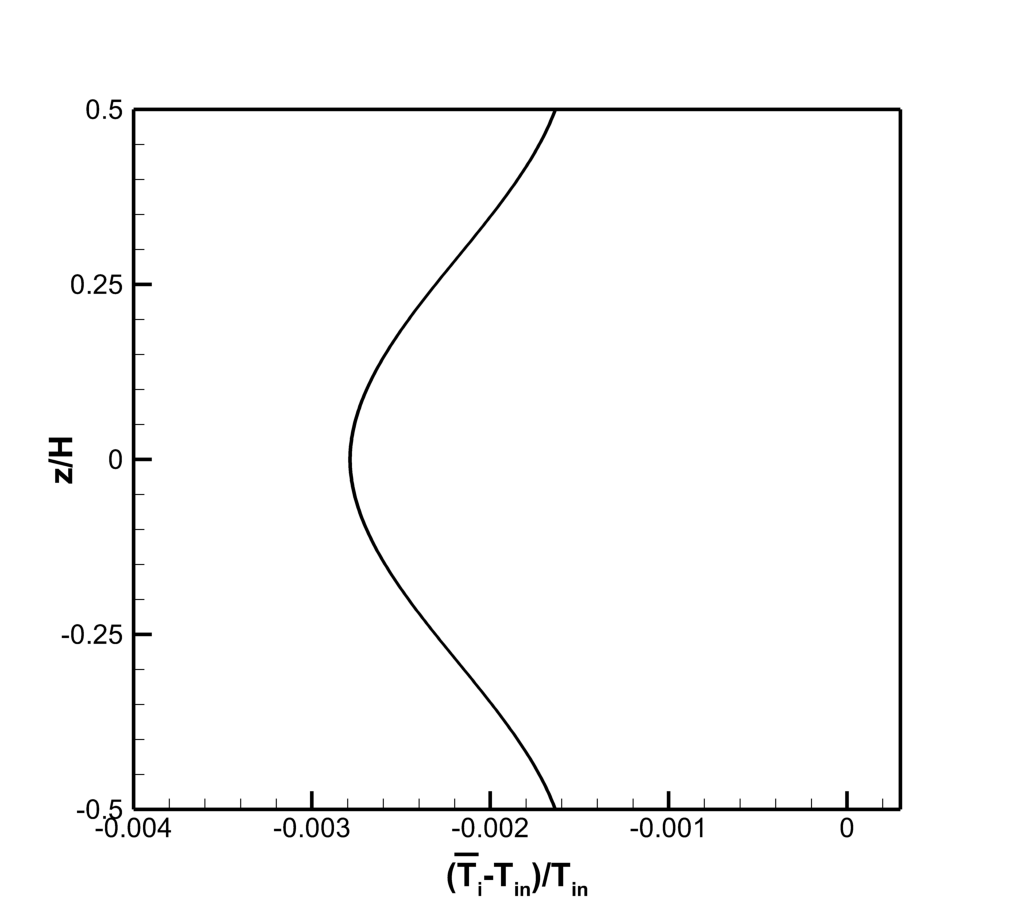}
\caption{}
\end{subfigure}
\caption{Flow properties in the settling chamber obtained from the RANS computation with the experimental \textmd{$P_\iin$} and \textmd{$H_\iin$} values entered as inflow. (a) Axial velocity \textmd{$\meant{u}_x$} in physical units. (b) Mach number $\meant{\Mach}$. (c) Static pressure $\meant{p}$ normalized upon \textmd{$P_\iin$}. (d) Static temperature $\meant{T}$ normalized upon \textmd{$T_\iin$}.}
\label{fig:profiles}
\end{figure}

\alert{The} velocity fluctuation need to be reconstructed by merging the experimental data with the numerical constraints. In effect, the numerical model is not able to take as inflow the experimental profiles provided by the measurements. The reason is attributed to the fact that the inflow frontier is too close to the nozzle. As a consequence a procedure has been applied to generate a set of inflow data admissible for the numerical model. First a calculation is carried out with uniform stagnation \alert{pressure $P_\iin$, temperature $T_\iin$, and enthalpy $H_\iin=c_p T_\iin$} as provided by the experimental tests. From this simulation the time-averaged profiles of velocity $\meant{u}_x$, Mach number $\meant{\Mach}$, static pressure $\meant{p}$ and temperature $\meant{T}$ are obtained. These profiles are shown in~\fref{fig:profiles}. 

\alert{From the experimental data,} the \alert{distribution of} the inflow velocity fluctuations $U_i$, \alert{computed for each random dimension ($i=1,2,...6)$}, is then obtained as
\begin{equation}
U_i = \meant{u}_{x,i}+\sigma_i\boldsymbol{\xi}
\end{equation} 
where \alert{$\boldsymbol{\xi}$ is made up of} either Gauss-Hermite points or sampled according to the  normal distribution $\npdf_0$ if one uses either the projection approach of \sref{Projection approach} or the regression approach of \sref{Regression approach}, respectively, and $\sigma_i=\mu_i\CV_i$ where $\CV_i$ is taken from the experiments; see \tref{inflow velocity uncertainties}. 

Next, these velocity fluctuations are used \alert{to compute $k$ velocity profiles $U^\circ_k(z)$ appropriate for the numerical simulations ($k=Q$ or $k=N$ based on the approach used, projection or regression) and to update individual Mach number profiles $M^\circ(z)$ as follows}
\begin{equation}
\Mach^\circ(z) = \dfrac{U^\circ(z)}{\sqrt{\dfrac{\gamma R T_\iin}{1+\dfrac{\gamma -1}{2} \meant{\Mach}(z)^2}}}\,.
\end{equation}
Subsequently the stagnation \alert{conditions} of pressure and enthalpy are provided by:
\begin{equation}
p^{\circ}_\iin(z) = \meant{p}(z)\left( 1 + \dfrac{\gamma - 1}{2} M^\circ(z)^2\right)^{\dfrac{\gamma}{\gamma-1}}\,,
\end{equation}
and
\begin{equation}
H^{\circ}_\iin(z) = c_{p}\left( 1 + \dfrac{\gamma - 1}{2} M^\circ(z)^2\right)\meant{T}(z)\,,
\end{equation}
where $\gamma=c_p/c_v$ is the ratio of specific heats, and $R=c_p-c_v$ is the specific gas constant. 

A parallel multi-threading scheme using \texttt{python} is implemented to perform several simulations for different inflow velocities simultaneously, in order to obtain either the $Q$ (at the Gauss-Hermite quadrature points), or $N$ (at randomly selected sample points) RANS evaluations needed by these approaches. 

\subsection{Numerical results}

\begin{figure}[H]
\minipage{0.5\textwidth}
  \includegraphics[width=0.95\linewidth]{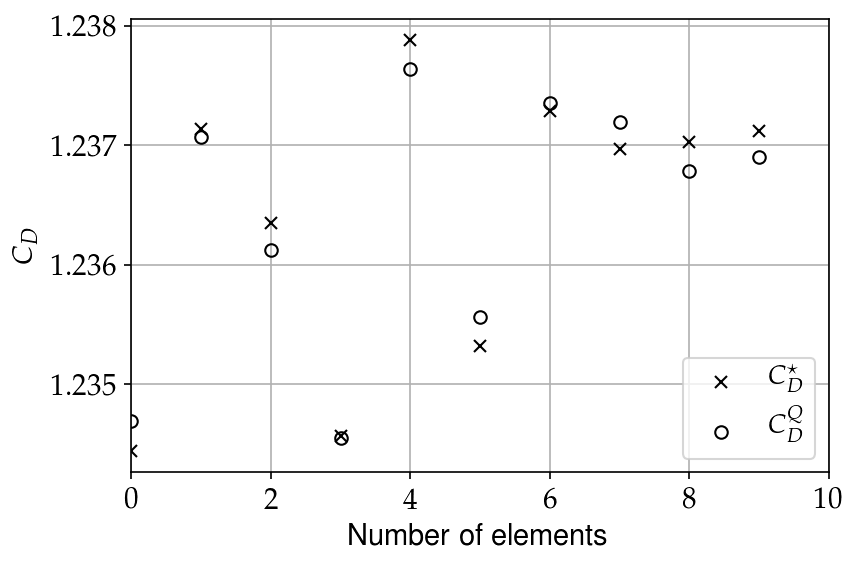} 
  \subcaption{$10$ samples}
  
\endminipage\hfill
\minipage{0.5\textwidth}
  \includegraphics[width=0.95\linewidth]{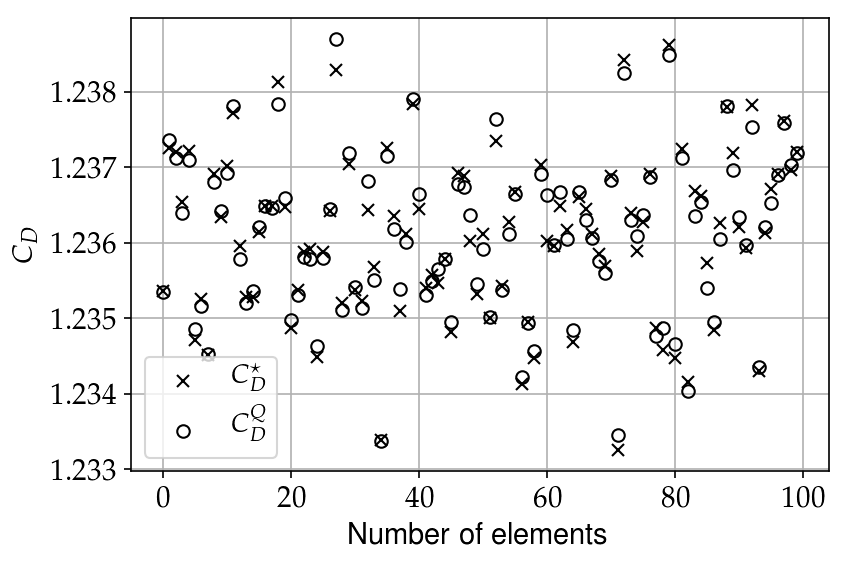} 
  \subcaption{$100$ samples}
  
 \endminipage\hfill
\minipage{0.5\textwidth}
  \includegraphics[width=0.95\linewidth]{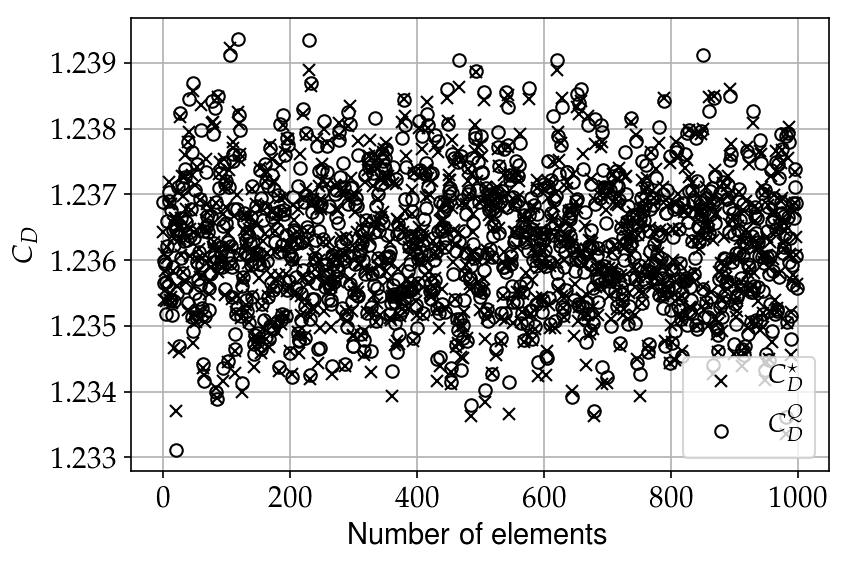}  
  \subcaption{$1,000$ samples}
   
 \endminipage\hfill
\minipage{0.5\textwidth}
  \includegraphics[width=0.95\linewidth]{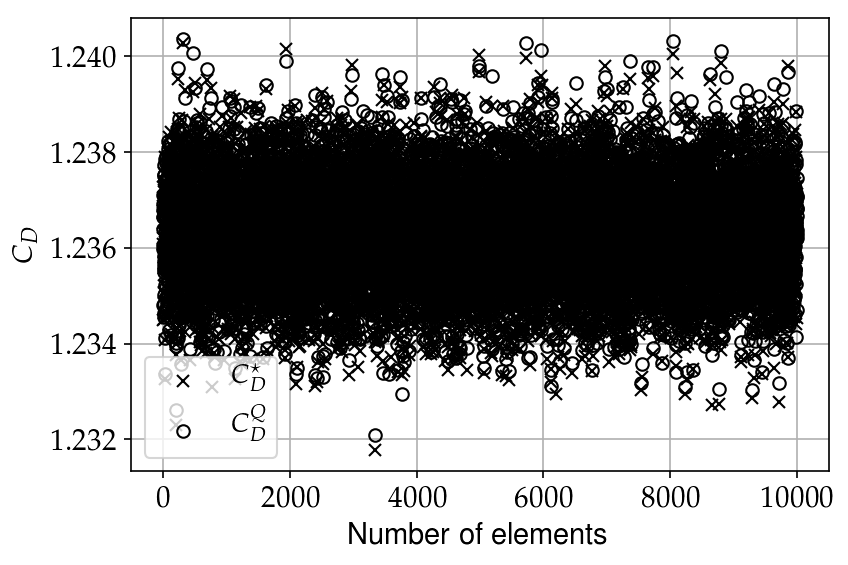}    
  \subcaption{$10,000$ samples}
 
\endminipage
\caption{Drag coefficient $C_D$ obtained by a polynomial chaos expansion using the projection (GQ) and regression (CS) approaches.}\label{CD measurements using the two approaches}
\end{figure}

Once all PCE coefficients have been computed, the two approaches can be compared in terms of the second-order statistics of the QoI $\QoI\equiv C_D$, using \eref{average_coeff} for its average and \eref{sd_coeff} for its standard deviation. These results are gathered in \tref{comparison methods}, the projection approach being considered as the reference (and most expensive) solution. \tref{comparison methods} shows a good agreement between the statistical moments computed using the coefficients obtained by the two approaches, highlighting an important physical aspect: a $1$ \% uncertainty in the inflow velocity (see \fref{Statistical moments for M=0.8}) results in a $0.09$ \% uncertainty in the drag coefficient. Therefore, in this way, the aim of this work has been reached. Besides, having computed the PCE coefficients $\boldsymbol{c}^Q$ and \alert{$\boldsymbol{c}^\star$}, the aerodynamic coefficient $C_D$ can be evaluated for any inflow velocity data $\xiv$ by:
\begin{equation}
C_D^Q(\xiv)\simeq\sum_{j=0}^\porder c^Q_j\Psi_j(\xiv)
\end{equation}
from the projection approach, or:
\begin{equation}
C_D^\star(\xiv)\simeq\sum_{j=0}^\porder \alert{c^\star_j}\Psi_j(\xiv)
\end{equation}
from the regression approach. Thus, generalized polynomial chaos allows us to obtain a surrogate model through which one can simulate, rapidly, the response of the cylinder to uncertainties at the wind tunnel inflow. Indeed, it is possible to compute the drag coefficient for different numbers of samples in order to highlight the behavior of the regression and projection approach, as done in \fref{CD measurements using the two approaches} for $10$ to $10,000$ samples of $\xiv$ with the PDF $\npdf_0$.

\begin{table}[H]
\begin{center}
\begin{tabular}{|c|c|c|}
\hline
$-$ & Regression approach	& Projection approach \\
\hline
$\mu_\porder$	&	1.236		&	1.236 \\
\hline
$\sigma_\porder$	& 0.001	&	 0.001\\
\hline
\nicola{$\CV$} (\%)	&	0.09\%	&	0.09\%  \\
\hline
\end{tabular}
\end{center}
\caption{Second-order statistical moments of $C_D$ computed from the expansion coefficients obtained through the projection and regression approaches.}\label{comparison methods}
\end{table}

\begin{figure}[H]
\begin{center}
\includegraphics[width=0.65\linewidth]{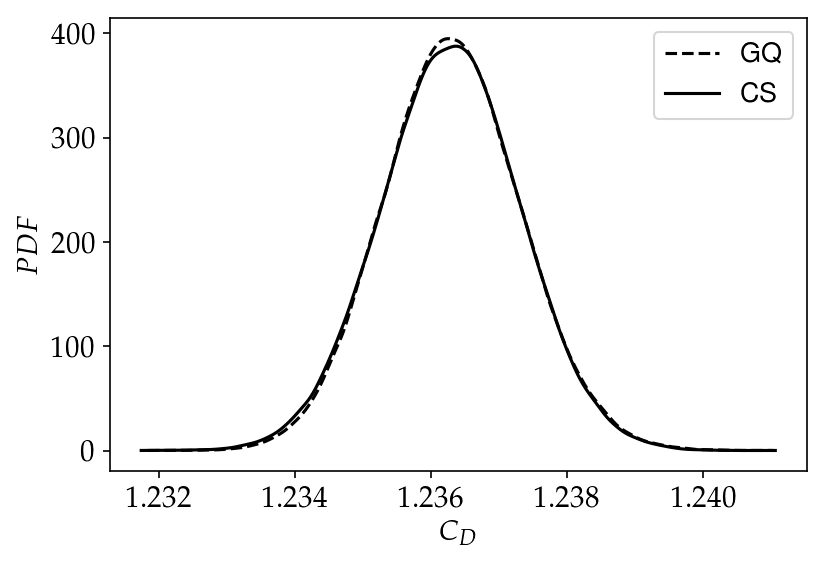}
\caption{Kernel density estimation of the PDF of $C_D$: comparison of the results obtained with the projection (GQ) and regression (CS) approaches using $100,000$ samples.}\label{CD kernel density estimation}
\end{center}
\end{figure}

Certainly, CS is able to reproduce with a good accuracy the distribution of a certain QoI as the number of samples with which the PCE is used is increased. In order to stress this observation, the PDFs of $C_D$ obtained by the kernel density estimation \bibrev{\cite{WAN95}} with both approaches and using a reasonable number of samples, can be compared in \fref{CD kernel density estimation}.

As a conclusion, one can notice from \fref{CD kernel density estimation} the good agreement between the projection approach and the regression approach remarking that the $\ell_{1}$-minimization of \eref{cjCS} offers an efficient method to construct the effective distribution of the QoI relying on its sparsity in a carefully chosen polynomial chaos basis. In particular, using the KL divergence (\ref{eq:KLdiv}) \cite{kullback1951information}, it can be shown that the drag coefficient follows a normal distribution, as the random fluctuations of the inflow velocities.

\subsection{Sensitivity analysis}

In order to further characterize the influence of the random inflow parameters, a sensitivity analysis can be performed that quantifies the respective effects of each input variable (or combinations thereof) onto the variance of the response of the drag coefficient. To do so, the Sobol indices have received much attention: each Sobol index $S_{d_1 d_2\dots d_s}$ is a sensitivity measure that describes which amount of the total variance is due to the uncertainties in a subset of $s$ input parameters. Denoting by $\mathcal{I}_d$ the set of indices corresponding to the polynomials of the basis depending only on the $d$-th variable parameter $\xi_d$, the main-effect PC-based Sobol indices are given by (see \emph{e.g.} \cite{sudret2008global}):
\begin{equation}
S_{d}=\dfrac{1}{\sigma_\porder^2}\sum_{j\in \mathcal{I}_d}c_j^2\,,
\end{equation}
owing to the normalization condition of \eref{eq:DDorthoPol}, with $\sigma^2_\porder=\sum_{j=1}^\porder c_j^2$ the variance of the QoI, \eref{sd_coeff}. More generally, if $\mathcal{I}_{d_1 d_2\dots d_s}$ is the set of indices corresponding to the polynomials of the basis depending only on the parameters $\xi_{d_1},\xi_{d_2},\dots\xi_{d_{s}}$, the $s$-fold joint sensitivity indices are:
\begin{equation}
S_{d_1 d_2\dots d_s}=\dfrac{1}{\sigma^2_\porder}\sum_{j\in \mathcal{I}_{d_1 d_2\dots d_s}}c_j^2\,.
\end{equation}
The main-effect and some joint ($2$-fold) Sobol indices computed for the six random input parameters are gathered in \tref{Main effect sensitivity indices by GQ and l1} and \tref{Joint sensitivity indices by GQ and l1} for the projection approach and for the regression approach. 

From these Sobol indices one can notice that the velocity fluctuations $U_3$ and $U_4$ influence the most the cylinder drag. The uncertainty on $U_{1..6}$ is almost uniform (see \tref{inflow velocity uncertainties}). The larger influence of the third and fourth velocity fluctuations hence seemingly results from their central position, while the uncertainties at the lower and upper sides of the settling chamber, described by the positions 1, 2 and 5, 6, have much less impact. Thus, $U_3$ and $U_4$ are the most sensitive parameters if one wants to interact with the transonic flow around the cylinder. Interestingly this indicates that flow defects further away from the central flow could be more acceptable. 

It must be eventually remarked that the low value of joint sensitivities indicate that mostly polynomial of order 1 are implicated in the surrogate model. An interesting consequence in that case is that the Gaussian \alert{inputs} naturally \alert{yield} a Gaussian output, as obtained here.

\begin{table}[H]
\begin{center}
\begin{tabular}{|c|c|c|c|c|c|c|}
\hline
$C_D$ & $S_{d_{1}}$ & $S_{d_{2}}$ & $S_{d_{3}}$ & $S_{d_{4}}$ & $S_{d_{5}}$ & $S_{d_{6}}$ \\
\hline
Projection & 0.0057 & 0.0027 & 0.4383 & 0.4439 & 0.0019 & 0.0166\\
\hline
Regression & 0.0064 & 0.0023 & 0.6982 & 0.3789 & 0.0009 & 0.0097\\
\hline
\end{tabular}
\end{center}
\caption{Main effect sensitivity indices of the inflow velocity parameters computed by Gauss-Hermite quadrature rule and $\ell_{1}$-minimization.}\label{Main effect sensitivity indices by GQ and l1}
\end{table}

\begin{table}[H]
\begin{center}
\begin{tabular}{|c|c|c|c|}
\hline
$C_D$ & $S_{d_{1}d_{2}}$ & $S_{d_{3}d_{4}}$ & $S_{d_{5}d_{6}}$ \\
\hline
projection & 4e-6 & 4e-4 & 7e-5 \\
\hline
regression & 2e-6 & 6e-4 & 9e-5 \\
\hline
\end{tabular}
\end{center}
\caption{Joint sensitivity indices of the inflow velocity parameters computed by Gauss-Hermite quadrature rule and $\ell_{1}$-minimization.}\label{Joint sensitivity indices by GQ and l1}
\end{table}

\section{Summary and conclusions}\label{sec:CL}

In this work a method to simulate a wind tunnel experiment has been studied, using CFD simulations and developing a polynomial surrogate model based on a polynomial chaos expansion (PCE) to account for the uncertainty of the inflow produced by the wind tunnel. In the first part of the work the inflow data variability and the flow around a cylinder in the wind tunnel test section have been analyzed in order to quantify the inflow uncertainty and validate the numerical model. In the second part the numerical wind tunnel has been set up. Unsteady Reynolds-averaged Navier-Stokes simulations have been performed and validated against experimental data. In the last part of this work, a stochastic approach has been developed to address the influence of parametric uncertainties on the numerical results. We have outlined two methodologies to construct the polynomial surrogate model: the projection approach and the regression approach. The first one has been used to built a reference solution based on the Gauss-Hermite quadrature rule in order to validate the second method. The latter, based on compressed sensing (CS) theory, relies on a so-called $\ell_{1}$-minimization and uses the concept of sparsity. The comparison between the two approaches highlights the good performances of CS, enhancing a method able to reproduce a certain quantity of interest with a low number of measurements or numerical simulations. After having obtained the surrogate model, the statistical distribution of the cylinder drag has been computed, simulating the inflow parameter variability. The cylinder drag remains little influenced by the inflow variations, and the central part of the flow is found to be the most influential. This result tends to minimize the problem of inflow variability on the quality of the numerical simulations, at least for this transonic cylinder case. 

In the future, improvements of such method will need to be \alert{carried} out to further assess the influence of wind tunnel flow quality. First in the present study the inflow uncertainty has only been accounted for in a reduced format to relax the computational cost of the numerous simulations. The next step would be to consider unsteady simulations, which were found to better match the experimental results and then to remove the constraint of two-dimensional flow in the simulation so as to evaluate the effect of the flow variability in the transverse direction. One important question is also that of the dependency between the inputs. In the present study they are considered as independent. Progress would be ensured by considering as dependent uncertain inputs in the form of modes of the inflow velocity rather than isolated data points. Such a modal decomposition would require two-point correlations of the flow field in the settling chamber. Furthermore it would certainly be interesting to open the list of uncertainties to geometrical features. High resolution laser scans as performed by~\cite{szoke2020developing} could offer useful data to envisage this problem.

\bibliography{biblio-v4}

\end{document}